\documentclass[amsmath,amssymb,prb]{revtex4}
\usepackage{graphicx}
\newcommand{\difd}{\mathrm{d}}
\newcommand{\Difd}{\mathrm{D}}

\newcommand{\bpp}{\psi}
\newcommand{\bee}{\eta}
\newcommand{\Tr}{\mathrm{Tr}}
\newcommand{\ham}{\mathcal{H}}
\newcommand{\oss}{\mathcal{O}_{\mbox{$s$SC}}}
\newcommand{\ocdw}{\mathcal{O}_{\mbox{CDW}}}

\newcommand{\opf}{\sqrt{2\pi}}

\begin{document}

\title{Functional renormalization group with counterterms for competing order parameters}
\author{Matthias Ossadnik$^{1,2}$, Carsten Honerkamp$^{2,3}$}
\affiliation{$^{1}$ Theoretical Physics, ETH H\"onggerberg, CH-8093 Z\"urich, Switzerland, \\
$^{2}$ Theoretical Physics, University of W\"urzburg, D-97074 W\"urzburg, Germany 
 \\
$^{2}$ Institute for Theoretical Physics C, RWTH Aachen University, D-52062 Aachen, Germany 
}
\date{November 22, 2007}

\begin{abstract}
We extend the fermionic functional renormalization group (fRG) flow for symmetry-broken states to a case of competing ordering tendencies in a one-dimensional model with attractive interactions. First, as an introduction to the fRG method with symmetry-breaking counterterms we study the reduced BCS model. 
Then the ground state phase diagram of an one-dimensional model with competing orders is analyzed by fRG and compared with bosonization.  We present a self-consistency argument which allows us to find the optimal counterterm for general models that cannot be solved exactly by mean-field theory. 
We show that the fRG is capable of reproducing the quantum phase transition between dominant pairing correlations and CDW order in a meaningful way.
\end{abstract}

\maketitle

\section{Introduction}

 Many of the phenomena in condensed matter physics can be understood as consequences of electronic correlations. 
Superconductivity, magnetism, and the quantum Hall effects are just a few examples of how collective behavior of electrons leads to macroscopic effects. In many cases the main effects can be ascribed to some kind of longer-range order. It is thus an important task of condensed matter theory to shed light on the question why and how electrons order in different ways. If the order parameter is known or restricted to a few possibilities, much progress can been made within mean-field approaches like the BCS theory of superconductivity. On the other hand, the question if and how the electrons order in microscopic models such as the Hubbard model is still not resolved in general. One reason for this question still being open is that various ordering tendencies strongly influence and compete with each other. 

The onset of long-range order is usually connected with a phase transition. One of the most powerful tools for the investigation of  phase transitions is the renormalization group (RG), laid out in its most common form by Wilson in 1971 \cite{wilson}. Wilson's RG is formulated in terms of the partition function of the system, that is, a functional integral over the degrees of freedom of the system. 
The idea is to integrate out the degrees of freedom one after another depending on their wavelength, in contrast to perturbation theory, where all modes are treated simultaneously. One begins at high energies and short wavelengths and works one's way down to low energies, taking into account the modes that are integrated out through changes in the parameters of the action of the remaining modes. This process is called renormalization, and the change of the parameters from high to low energies is called the RG flow.

Later, Polchinski \cite{polchinski} formulated the renormalization group as an exact flow equation for the partition function of the theory. In the following, this was generalized to other generating functionals, such as the one for one-particle irreducible (1PI) vertices that will be used in this work. The flow equations for this generating functional were first derived for scalar fields by Wetterich \cite{wetterich}, and later for fermionic systems\cite{tnt,goettingen,kopietz}. By now it has been widely used to investigate many-fermion systems in various contexts (e.g. Refs. \onlinecite{zanchi,halboth,rice,katanin,tempflow,wang}. 
In these systems the dominant question is which type of symmetry-breaking takes place at the transition, while the precise nature of the transition is often only of secondary interest. 
For the RG schemes used in this context, the term {\em functional renormalization group}  applies to the method in its present form in a twofold way: one uses a flow equation for a generating functional of fermionic fields to obtain flow equations for vertices, which are in turn functions of the generalized quantum numbers (wavevectors, Wannier orbitals) of the degrees of freedom. This has to be contrasted with usual RGs which are in general flow equations for just a small number of coupling constants.

The functional RG was in particular employed to study instabilities of the Fermi surface of the 2D Hubbard model towards different symmetry broken states such as antiferromagnetism, superconductivity, and ferromagnetism by various researchers (e.g. Refs. \onlinecite{zanchi,halboth,rice,katanin,tempflow}). However, for technical reasons, it was not possible in these studies to obtain quantitative results for physical quantities like the order parameter, as the flows used were restricted to the symmetric phase. Instabilities in these studies would only show up in the form of diverging susceptibilities. In 2004, Salmhofer et al. \cite{lauscher} found a way to continue the RG flow into phases with broken symmetry by including a small external field in the model, which is explained in detail in section \ref {sec:rbcs}.
 As these external fields bias the results, they are by no means an optimal method for calculating order parameters. However, Gersch et al. \cite{gersch} extended this idea by replacing the unphysical external field by a counterterm, an external field that is added to the interacting Hamiltonian but at the same time subtracted from the free Hamiltonian, so that - at least in principle - the calculation is not biased any more (as long as calculations are performed exactly). This will be discussed in section \ref{ch:counterterms}.

Until now, the method of counterterms has only been applied to a model where the flow equations can be solved exactly\cite{gersch}. The goal here is to see how the various approximations involved when dealing with non-trivial systems affect the method. More concretely, we want to investigate if the counterterm scheme {\em a)} is able to deal with situations where mean-field theory is not exact and {\em b)} in situations in which instabilities occur in more that one channel at a time. 
We consider the attractive Tomonaga-Luttinger model at half-filling as a simple toy model. This model allows numerically cheap computations, whilst at the same time exhibiting interesting behaviour with two competing instabilities in the region of the phase diagram considered. A further advantage is that very much is known about the Tomonaga-Luttinger model (as will be discussed in section \ref{sec:tomonaga}), so that the data obtained by the functional RG approach can be compared with existing information. 

 Of course, as the Tomonaga-Luttinger model is already well-understood, no new physical results will be presented. Our interest here is purely methodological, and is thought to serve as a preliminary, but important test of the method before applying it to more interesting situations, such as the repulsive 2D Hubbard model. Another test in the attractive 2D Hubbard model produced promising results that this approach can be used to get beyond standard mean-field calculations\cite{gersch2d}.
 
Before embarking onto the derivation of the necessary fermionic functional RG equations, let us comment on an alternative promising approach for flows into symmetry-broken regimes in interacting many-fermion systems. Instead of following the four-fermions interactions down to low scales, it is also possible to decouple the fermions via an adaptive Hubbard-Stratonovitch transformation in the relevant bosonic channels (which channels one should actually choose can be read off from the fermionic flows in the symmetric phase)\cite{bosonizerefs}. This has the advantage that the large fermionic interactions are replaced by nearly massless bosons and can cause no further problems in the truncated flow equations. Furthermore collective fluctuations of the order parameter (i.e. the ordering boson) can be treated more easily. Often the connection to standard models of statistical mechanics such as $O(N)$-models is feasible. On the back side, the generation of couplings in channels not present in the initial interaction (such as $d$-wave pairing in the 2D Hubbard model on the square lattice) and the removal of regenerated fermionic interactions are technically challenging\cite{gies}, and the bosonic propagators and the Yukawa couplings to the fermions need to be approximated strongly for practical calculations. This in turn reintroduces some degree of bias and complicates the application to more realistic models. While these problems can certainly be taken care of to some degree, we feel that in addition one should try to develop a complementary,  purely fermionic approach that is able to capture the most relevant corrections beyond the mean-field studies and that could be used to determine the basic gap structures of unconventional superconductors in a relatively straightforward setup.

\section{The Functional Renormalization Group for the 1PI Vertices}
\subsection{Derivation of the Flow Equations} \label{sec:floweq}
In this section the fRG flow equations for the one-particle irreducible (1PI) vertex functions are described. The derivation presented here follows Ref. \onlinecite{tnt}. For alternative derivations, see, e.g., Ref. \onlinecite{kopietz}.
The fRG is based on the evaluation of a generating functional describing a quantum many-body system by introducing a parameter $\chi$ into the functional such that for some value $\chi_i$ the functional can be evaluated exactly and for some other value $\chi_f$ the functional describes the system one is interested in. 
This parameter is called the flow parameter. There are many different ways of impleneting this idea, which correspond to different choices of generating functionals and flow parameters. Evaluating the functional is then done by virtue of the identity
\begin{equation}
\lim_{\chi\rightarrow\chi_f}f(\chi) = f(\chi_i) + \int_{\chi_i}^{\chi} \dot{f}(\tilde{\chi}) \, \difd\tilde{\chi}.
\end{equation}
The limit in the above formula will turn out to be important, as generating functionals in the thermodynamic limit may possess singularities, so the result in general depends on how $\chi_f$ is approached. This corresponds to the fact that in the thermodynamic limit phase transitions may occur.

The starting point for the derivation is the partition function - the generating functional of correlation functions. For fermionic systems it is given by the Grassmann functional integral
\begin{equation}
Z[\eta] = \frac{1}{Z_0}\int\Difd\psi \, e^{-S[\psi] - (\eta,\psi)},
\end{equation}
where we have have also inserted a normalization factor and Grassmannian source fields. The action is now separated into a free (quadratic) part and an interacting part,
\begin{equation}
S[\psi] = \frac{1}{2}(\psi,Q\psi) + V[\psi],
\end{equation}
where the $\psi$ are Grassmannian fields, which are viewed as Grassmann algebra-valued vectors indexed by the multiindex $X \equiv (x,\tau, c,s,\ldots)$. The charge index $c$ is introduced to write both $\bar{\psi}$ and $\psi$ as components of one field. $Q$ is the inverse free propagator of the system, which is a matrix in the vector space generated by $X$. With the introduction of the charge index for the fields, $Q$ is antisymmetric. The notation $(\cdotp,\cdotp)$ denotes the standard inner product of the vector space generated by the fields $\psi_X$. The normalization factor $Z_0$ with these conventions is given by the integral over the free action only, $Z_0 = \sqrt{\det Q}$. The square root comes from using the charge index. 

 Next we introduce the real flow parameter $\chi$ into the free action, $Q\rightarrow Q_\chi$, where the exact form of the parameter dependence of $Q_\chi$ need not be specified for the derivation. This induces a flow parameter dependence of the partition function as well, $Z\rightarrow Z_\chi$. In order to arrive at the desired flow equations for the generating functional of 1PI vertices, which is also called the effective action, we have to perform a series of transformations on $Z_\chi$.

First, we define the generating functional of connected correlation functions by
\begin{equation}
W_\chi[\eta] = \log\left\{\int \frac{D(\psi)}{\sqrt{\det Q_\chi}} e^{-S_\chi[\psi]  - (\eta,\psi)}\right\}.
\label{defW}
\end{equation}
We define the field expectation values as
\begin{equation}
\phi :=  -\frac{\partial W_\chi}{\partial\eta}[\eta].
\label{defphi}
\end{equation}
This naturally induces the inverse $\eta[\phi] = \eta_\chi [\phi]$. Equipped with these definitions one can perform the Legendre transformation of $W_\chi[\eta]$ with respect to $\eta$ to arrive at the effective action
\begin{equation}
\Gamma_\chi[\phi] = -W_\chi\left[\eta [\phi]\right] - \left(\phi, \eta[\phi]\right).
\label{defgamma}
\end{equation}
In order to derive a flow equation for $\Gamma_\chi[\phi]$, we differentiate eq.(\ref{defgamma})with respect to $\chi$ to find
\begin{equation}
\dot{\Gamma}_\chi = -\partial_\chi W - \left(\dot{\eta},\partial_\eta W\right) - \left(\phi,\dot{\eta}\right),
\label{gammadot}
\end{equation}
where all dependences on $\phi$ and $\eta$ have been omitted and the dot denotes the total derivative with respect to $\chi$. On inserting the definition of $\phi$ given in Eq.\ref{defphi}, one finds that all terms except the first one in Eq. \ref{gammadot} cancel, so that
\begin{equation}
\dot{\Gamma}_\chi\left[\phi\right] = -\partial_\chi W_\chi\left[\eta\left[\phi\right]\right].
\end{equation}
Hence, it is clear that it is sufficient to evaluate the derivative of $W_\chi$, which is readily done by taking into account the definition of $W_\chi$ in eq.(\ref{defW}). Writing $Z_\chi[\eta] = \exp\left\{W_\chi[\eta]\right\}$ on has
\begin{equation}
\partial_\chi Z_\chi = \partial_\chi e^{W_\chi} = \left(\partial_\chi W_\chi\right) Z_\chi 
\end{equation}
and (omitting some subscripts $\chi$)
\begin{eqnarray}
\nonumber \partial_\chi Z &=& \sqrt{\det Q}\left(\partial_\chi\frac{1}{\sqrt{\det Q}}\right) Z - \frac{1}{2}\int\frac{\Difd(\bpp)}{\sqrt{\det Q}}  \dot{Q}_{ij}\partial_{Q_{ij}} e^{-S[\bpp] + (\eta,\psi)}\\ \nonumber
&=& \sqrt{\det Q}\left(\partial_\chi\frac{1}{\sqrt{\det Q}}\right) Z - \frac{1}{2}\int\frac{\Difd(\bpp)}{\sqrt{\det Q}} \left(\psi,\dot{Q}\psi\right) e^{-S[\bpp] + (\eta,\psi)} \\ \nonumber
&=& \sqrt{\det Q}\left(\partial_\chi\frac{1}{\sqrt{\det Q}}\right) Z - \left(\delta_\eta,\dot{Q}\delta_\eta\right)\frac{1}{2}\int\frac{\Difd(\bpp)}{\sqrt{\det Q}}  e^{-S[\bpp]  + (\eta,\psi)}\\ \nonumber
&=& \left\{\sqrt{\det Q}\left(\partial_\chi\frac{1}{\sqrt{\det Q}}\right) - \frac{1}{2}\left(\delta_\eta,\dot{Q}\delta_\eta\right)\right\}Z. 
\label{derivdotZ}
\end{eqnarray}
The first term still has to be evaluated, which is readily done employing the well known identity from linear algebra, $\ln \det = \Tr\ln$, to obtain
\begin{eqnarray}
\partial_\chi\frac{1}{\sqrt{\det Q}} &=& \partial_\chi \exp\left\{-\frac{1}{2}\Tr\ln Q\right\} \nonumber \\
&=& -\frac{1}{2}\Tr \dot{Q}Q^{-1}\frac{1}{\sqrt{\det Q}}. 
\end{eqnarray}
Finally, one has to employ the definition of $Z = e^W$ into eq.(\ref{derivdotZ}) and evaluate the derivatives acting on $Z$ in the second term of eq.(\ref{derivdotZ}), which leads to
\begin{eqnarray}
\left(\delta_\eta,\dot{Q}\delta_\eta\right)Z &=& \left(\delta_\eta,\dot{Q}\delta_\eta\right)e^W \nonumber\\
&=& \left(\left(\delta_\eta W,\dot{Q}  \delta_\eta W\right) + \Tr\left(\dot Q \delta_\eta\delta_\eta W\right)\right) Z,
\end{eqnarray}
where $\delta_\eta\delta_\eta W$ is understood as a matrix in the vector space indexed by the usual multiindex, where the indices of the sources $\bee$ define the matrix indices. Putting all together, we have arrived at
\begin{equation}
\partial_\chi W = -\frac{1}{2}\left\{\Tr\left(\dot{Q}Q^{-1}\right) + \left(\delta_\eta W, \dot{Q}\delta_\eta W\right) + \Tr \left(\dot{Q}\delta_\eta\delta_\eta W\right)\right\},
\end{equation}
which now has to be translated into a formula for $\dot{\Gamma}$ by replacing $(\bee)$ by $(\bar{\phi},\phi)$ and $W$ by $\Gamma$. This is done using the identities (see, e.g. Ref. \onlinecite{peskin}) 
\begin{eqnarray}
\delta_\phi\Gamma &=& \eta, \nonumber  \\
\mathbf{1} &=& \left(\delta^2_\phi\Gamma\right)\left(\delta^2_\eta W\right).
\label{GWrelation}
\end{eqnarray}
Upon using them, one finds that
\begin{equation}
\dot{\Gamma} = \frac{1}{2}\Tr\left(\dot{Q}Q^{-1}\right) + \frac{1}{2}\left(\bar{\phi},\dot{Q}\phi\right) + \frac{1}{2}\Tr\left(\dot{Q}\tilde{\Gamma}^{-1}\right),
\label{gammafeq}
\end{equation}
where $\tilde{\Gamma}_{ij}$ denotes the matrix $\partial_{\phi_i}\partial_{\phi_j}\Gamma$. This is the central result of this section, obtained first in Ref. \onlinecite{tnt} . An equivalent formula for the scalar case was worked out  before in Ref. \onlinecite{wetterich}. The trace in the last term can be understood as a one-loop diagram, i.e. the flow can be evaluated quite simply for cases where one has a good ansatz for $\Gamma$. For fermions however, $\Gamma$ needs to be expressed in monomials of Grassmann fields which leads to a infinite set of coupled one-loop equations for the vertex functions.
 This will be explained in  the remainder of this section.
 
The expansion of  $\Gamma$ into a basis of monomials in the fields $\phi$ is done by writing
\begin{equation}
\Gamma[\phi] = \sum_{m\geq 0}\gamma^{(m)}[\phi]
\end{equation}
with
\begin{equation}
\gamma^{(m)}[\phi] = \frac{1}{m!}\int\difd^m\mathbf{X}\gamma_m(\mathbf{X})\phi^m(\mathbf{X}).
\end{equation}
Here $\mathbf{X} = (X_1,\cdots,X_m)$ is an m-tuple of multiindices, and $\phi^m(\mathbf{X}) = \phi(X_1)\cdots\phi(X_m)$. The $\gamma_m(\mathbf{X})$ are chosen to be totally antisymmetric, reflecting the fermionic nature of the fields involved. Furthermore, we similarly expand
\begin{equation}
\frac{\delta}{\delta\phi(X)}\frac{\delta}{\delta\phi(Y)}\Gamma(\phi) = \sum_{m\geq 2}\tilde{\gamma}^{(m)}(X,Y;\phi),
\end{equation}
where
\begin{equation}
\tilde{\gamma}^{(m)}(X,Y;\phi) = \frac{1}{(m-2)!}\int\difd^m\mathbf{X}'\gamma_m(X,Y,\mathbf{X}')\phi^{m-2}(\mathbf{X}')
\end{equation}
One finds in particular that $\tilde{\gamma}^{(2)}$ does not depend on $\phi$, so that
\begin{equation}
\frac{\delta^2\Gamma(\phi)}{\delta\phi(X)\delta\phi(Y)}\Bigg|_{\phi(X)=\phi(Y)=0} = \gamma_2(X,Y).
\end{equation}
From eq.(\ref{GWrelation}) it follows thus that
\begin{equation}
\gamma_2 = \mathcal{G}^{-1},
\end{equation}
where $\mathcal{G}$ denotes the full propagator, so that by defining
\begin{eqnarray}
\frac{\delta^2\Gamma(\phi)}{\delta\phi(X)\delta\phi(Y)} &=& \gamma_2(X,Y) + \tilde{\Gamma}(X,Y;\phi) \nonumber \\
&=& \gamma_2\left(1 + \mathcal{G}\tilde{\Gamma}(\phi)\right).
\end{eqnarray}
With this definitions, the flow equation, eq.(\ref{gammafeq}) reads
\begin{equation}
\frac{1}{2}\left\{\Tr\left( Q^{-1}\dot{Q}\right) + \left(\phi,\dot{Q}\phi\right) + \Tr\left[\mathcal{G}\dot{Q}\left(1+\mathcal{G}\tilde{\Gamma}(\phi)\right)^{-1}\right]\right\}.
\label{gamdotnonpol}
\end{equation}
This equation is still non-polynomial because of the last term. This term, however, may be written as a power series as
\begin{eqnarray}
\Tr\left[\mathcal{G}\dot{Q}\left(1+\mathcal{G}\tilde{\Gamma}(\phi)\right)^{-1}\right] &=& 
 \Tr\left(\mathcal{G}\dot{Q}\right) - \Tr\left(\mathcal{G}\dot{Q}\mathcal{G}\tilde{\Gamma}(\phi)\right) \nonumber \\
& & + \sum_{p\geq 2}(-1)^p \Tr\left[\mathcal{G}\dot{Q}\left(\mathcal{G}\tilde{\Gamma}(\phi)\right)^p\right].
\label{powerseries}
\end{eqnarray}
By defining the so-called \emph{single scale propagator} $S$ as
\begin{equation}
S = -\mathcal{G}\dot{Q}\mathcal{G},
\label{defsprop}
\end{equation}
and comparing equal powers of the fields in eq.(\ref{gamdotnonpol}) using eq.(\ref{powerseries}) we finally arrive at the flow equations for the 1PI vertices:
\begin{eqnarray}
\dot{\gamma}^{(2)}(\phi) &=& \frac{1}{2}\left(\phi,\dot{Q}\phi\right) + \frac{1}{2}\Tr\left(S\tilde{\gamma}^{(4)}(\phi)\right),\\
\dot{\gamma}^{(4)}(\phi) &=& \frac{1}{2}\Tr\left[S\tilde{\gamma}^{(6)}(\phi)\right] - \frac{1}{2}\Tr\left[S\tilde{\gamma}^{(4)}(\phi)\mathcal{G}\tilde{\gamma}^{(4)}(\phi)\right],\\
\dot{\gamma}^{(6)}(\phi) &=& \frac{1}{2}\Tr\left[S\tilde{\gamma}^{(8)}(\phi)\right] - \frac{1}{2}\Tr\left[S(\tilde{\gamma}^{(6)}\mathcal{G}\tilde{\gamma}^{(4)} + \tilde{\gamma}^{(4)}\mathcal{G}\tilde{\gamma}^{(6)}\right] + \frac{1}{2}\Tr\left[S\tilde{\gamma}^{(4)}\mathcal{G}\tilde{\gamma}^{(4)}\mathcal{G}\tilde{\gamma}^{(4)}\right], \\
\vdots \nonumber
\end{eqnarray}
Actually, one has an infinite hierarchy of ordinary differential equations, but as in actual computations only the lowest equations are used, the equations for the higher order vertices are not displayed here.

 In order to compare coefficients the RHS of the flow equations have to be antisymmetrized, because the LHS are. Carrying out this antisymmetrization one arrives at the following set of equations for $\dot{\gamma}_2$ and $\dot{\gamma}_4$:
\begin{eqnarray}
\dot{\gamma}_2(X_1,X_2) &=& \dot{Q}(X_1,X_2) + \frac{1}{2}\int\difd X_3\difd X_4 S(X_4,X_3)\gamma_4(X_1,X_2,X_3,X_4), \label{g2dot} \\
\dot{\gamma}_4(\mathbf{X}) &=& \frac{1}{2}\int\difd Y_1\difd Y_2\gamma_6(\mathbf{X},Y_1,Y_2)S(Y_2,Y_1) - \frac{1}{2}\int \difd^4\mathbf{Y} L(\mathbf{Y} B(\mathbf{X},\mathbf{Y}), \label{g4dot}
\end{eqnarray}
where $\mathbf{Y} = (Y_1,\cdots,Y_4)$ and similarly for $\mathbf{X}$, and furthermore
\begin{eqnarray}
L(\mathbf{Y}) &=& S(Y_1,Y_2)G(Y_3,Y_4) + S(Y_3,Y_4)G(Y_1,Y_2)\\
B(\mathbf{X},\mathbf{Y}) &=& \gamma_4(X_1,X_2,Y_2,Y_3)\gamma_4(Y_4,Y_1,X_3,X_4)  \nonumber \\
& & \gamma_4(X_1,X_3,Y_2,Y_3)\gamma_4(Y_4,Y_1,X_2,X_4)  \nonumber \\&&+\gamma_4(X_1,X_4,Y_2,Y_3)\gamma_4(Y_4,Y_1,X_2,X_3).
\end{eqnarray}

 It is thus obvious that one possibility to truncate the infinite hierarchy of flow equations is to neglect the contribution of $\gamma_6$ in eq.(\ref{g4dot}), and to solve only the closed system of equations for $\gamma_2$ and $\gamma_4$. The higher (non-1PI) correlation functions are then given by tree diagrams consisting of full propagators and the 1PI vertices given by the $\gamma_4$. It is this strategy that will be pursued in the following. A justification for dropping $\gamma_6$ under certain conditions is given in Ref. \onlinecite{tnt}.

\subsection{Flows into Phases with Broken Symmetry} \label{sec:rbcs}
The main focus of this work is the calculation of order parameters corresponding to states with spontaneous symmetry breaking (SSB). In the setting described above, the fRG can \emph{not} be used to calculate these order parameters. The technical reason for this is that if the initial system is invariant under some symmetry group $G$, then so will be all the 1PI vertices in the expansion of $\Gamma$. Now the RHS of the flow equations contain only sums of products of the 1PI vertices. But if all the vertices are invariant under $G$, then so are products of them. Thus no symmetry breaking terms can be generated during the flow. If the flow starts in the symmetric phase, it will remain there. 
In the case of a second order phase transition this will lead to divergences, because in this case susceptibilities diverge, and this divergence is reflected in the four-point correlation functions, as a susceptibility is renormalized by the four-point correlation function of the fermions.
This method has already been used extensively to study the 2D Hubbard model\cite{zanchi}, \cite{rice}, \cite{tempflow}. In this case, of course, the flow cannot be performed until the end, so that the low-energy behaviour of the system can not be extracted precisely. 

One workaround for this problem has been proposed in Ref. \onlinecite{lauscher} and consists of the application of a small external field that corresponds to the symmetry breaking one expects. For example, if one expects an instability towards $s$-wave superconductivity, one would add to the initial Hamiltonian an (infinitesimally small) $s$-wave pairing field $\Delta_{\mbox{ext}}$,
\begin{displaymath}
\Delta_{\mbox{ext}}\sum_{\vec{k}} \left(c^\dagger_{\vec{k}\uparrow}c^\dagger_{-\vec{k}\downarrow}\right) + \mbox{h.c.}
\end{displaymath}
In this approach, the symmetry is broken from the very beginning of the flow, so that 1PI vertices do exist that break $G$, thus allowing for a flow of other $G$-breaking terms, such as the order parameter. On the other hand, of course, results obtained with this method will depend on the magnitude of the applied field, and only in the limit of vanishing external field the desired unperturbed results are recovered. Of course this idea corresponds to the above mentioned fact that generating functionals in the thermodynamic limit possess singularities that mark phase transitions. The procedure of introducing a small external field is similar to choosing the correct ordering of the limits $N\rightarrow \infty$ and $\Delta\rightarrow 0$ in statistical physics.

The model considered in Ref. \onlinecite{lauscher} is the so-called reduced BCS (rBCS) model, which is described by a many-fermion system where the only interaction is via the Cooper channel with zero total momentum. Because there exists only one channel, mean-field theory is exact for this model. One can show that this implies that the truncation of the flow equations discussed in section \ref{sec:floweq} is also exact. This makes this model a good testing ground for studying the effect of an external field without having to care about large couplings, effects of the truncation and the like. In the remainder we briefly summarize the results of Ref. \onlinecite{lauscher}.

 Concretely, the model is described by the Hamiltonian
\begin{equation}
\ham = \ham_0 - \frac{v}{V} \sum_{\vec{k},\vec{p}} f(\vec{k})f(\vec{p})c^\dagger_{\vec{k}\downarrow}c^\dagger_{-\vec{k}\uparrow}c^\dagger_{\vec{p}\uparrow}c^\dagger_{-\vec{p} \downarrow},
\end{equation}
where $f(\vec{k})$ describes the angular momentum of the pairing and will for simplicity be set to $f(\vec{k})\equiv 1$ in the following, $v$ is the interaction strength and $V$ is the volume of the system. This choice of $f(\vec{k})$ corresponds to s-wave pairing and does not lead to a loss of generality, as in the rBCS model the different pairing channels decouple. 

Next, the external field is introduced as above by adding the term
\begin{equation}
\ham_{\mbox{ext}} = \Sigma \sum_{\vec{k}} c^\dagger_{\vec{k}\downarrow}c^\dagger_{-\vec{k}\uparrow} + \Sigma^\dagger\sum_{\vec{k}}c_{\vec{k}\uparrow}c_{-\vec{k}\downarrow},
\end{equation}
describing an external $s$-wave pairing field $\Sigma$, to the Hamiltonian. The symbol $\Sigma$ instead of the more common $\Delta$ is chosen here because the external field is treated as an initial condition for the self-energy as described below.

 In order to handle the new vertices (and lines) appearing due to the symmetry breaking, it is convenient to introduce the usual superconducting Nambu spinors (in the functional integral language, see Ref. \onlinecite{altland}) und to group wavevectors $\vec{k}$ and Matsubara frequencies $\omega=k_0$ to a multiindex $k$ ,
\begin{equation}
\Psi_k = \left(\begin{array}{c} \psi_{k\uparrow}\\\bar{\psi}_{-k\downarrow}\end{array}\right),\qquad
\bar{\Psi}_k = \left(\bar{\psi}_{k\uparrow}, \psi_{-k\downarrow}\right),
\end{equation}
so that the matrix representation of the quadratic part of the action becomes
\begin{equation}
Q = \left(\begin{array}{cc} Q^{(0)}_k & \Sigma \\ -\Sigma^\dagger & \left[Q^{(0)}_{-k}\right]^{\mbox{T}}\end{array}\right).
\end{equation}
Here $Q^{(0)}$ denotes the propagator in the symmetric phase. The external field is introduced as initial condition for the self-energy, so that $\Sigma$ in the definition of the inverse free propagator is actually zero, and appears only in the full propagator, connecting fields with opposite Matsubara frequencies.

Diagrammatically, the off-diagonal elements of the propagator can be viewed as new types of lines that create/annihilate two particles forming a singlet. During the flow, these new lines also lead to new vertices that do not conserve particle number. In the rBCS model, there are two of them corresponding to creation/annihilation of four particles forming two singlets. However, the anomalous terms creating particles are related to those annihilating particles by complex conjugation. Thus, choosing the external field to be real, their numerical values are actually identical. The anomalous vertex will be denoted by $w$ in the following. As this vertex is only generated during the flow, its initial condition is $w = 0$.

 Before stating the results it is important to mention that the flow equations derived in section \ref{sec:floweq} have to be modified in order to obtain the correct result for the gap. The Katanin modification\cite{kataninmod} of the truncated flow equations, eqns.(\ref{g2dot}) and (\ref{g4dot}), amounts to the substitution 
\begin{equation}
S\longrightarrow \dot{G}
\end{equation}
in the flow equation of $\gamma^{(4)}$, where the dot indicates the derivative with respect to the flow parameter. The advantage of this modification is that it improves the fulfillment of Ward identities, which are violated in the truncated fRG scheme. Katanin has shown that as a consequence the flow of symmetry breaking order parameters is also improved, so that in the ladder (or bubble) approximation, where only particle-particle (particle-hole) diagrams are kept, the mean-field equation for the order parameter is recovered, which is not the case when using the original truncated flow equations. The Katanin modification is thus a necessary ingredient when performing flows in symmetry broken phases and will be always used in the remainder of this work.

The (truncated) flow equations may be obtained from eqns. (\ref{g2dot}) and (\ref{g4dot}) and read
\begin{eqnarray}
\dot{v} &=& \dot{B}\left(v^2 + w^2\right) + 2 \dot{A}vw \\
\dot{w} &=& \dot{A}\left(v^2 + w^2\right) + 2\dot{B}vw\\
\dot{\Sigma} &=& C\left(v+w\right) \label{rbcsflow}
\end{eqnarray}
where $\dot{A},\dot{B}$ and $C$ refer to the loop integrals given by
\begin{eqnarray}
A &=& -T\sum_\omega\int dk \left(\frac{\Sigma\chi(s)^2}{|Q|^2 + (\Sigma\chi(s))^2}\right)^2 \\
B &=& T\sum_\omega\int dk \frac{(\omega^2 + \epsilon_k^2)\chi(s)^2}{\left(|Q|^2 + (\Sigma\chi(s))^2\right)^2}\\
C &=& T\sum_\omega\int dk \frac{2\chi(s)\dot{\chi(s)}\Sigma|Q|^2}{|Q|^2 + (\Sigma\chi(s))^2}.\label{rbcsloops}
\end{eqnarray}
Here we have only written the equation for the vertices with zero total incoming frequency. Vertices with nonzero total frequencies  also flow, but more weakly and do not give rise to any self-energy corrections in the thermodynamic limit due to the restricted wavevector-structure of the bare interaction. We could have equally well started with the doubly reduced interaction where only zero total frequency pairs interact. The anomalous selfenergy remains frequency independent.

\begin{figure}[h]
\label{rbcsresults}
\centering
\includegraphics[width = 10cm]{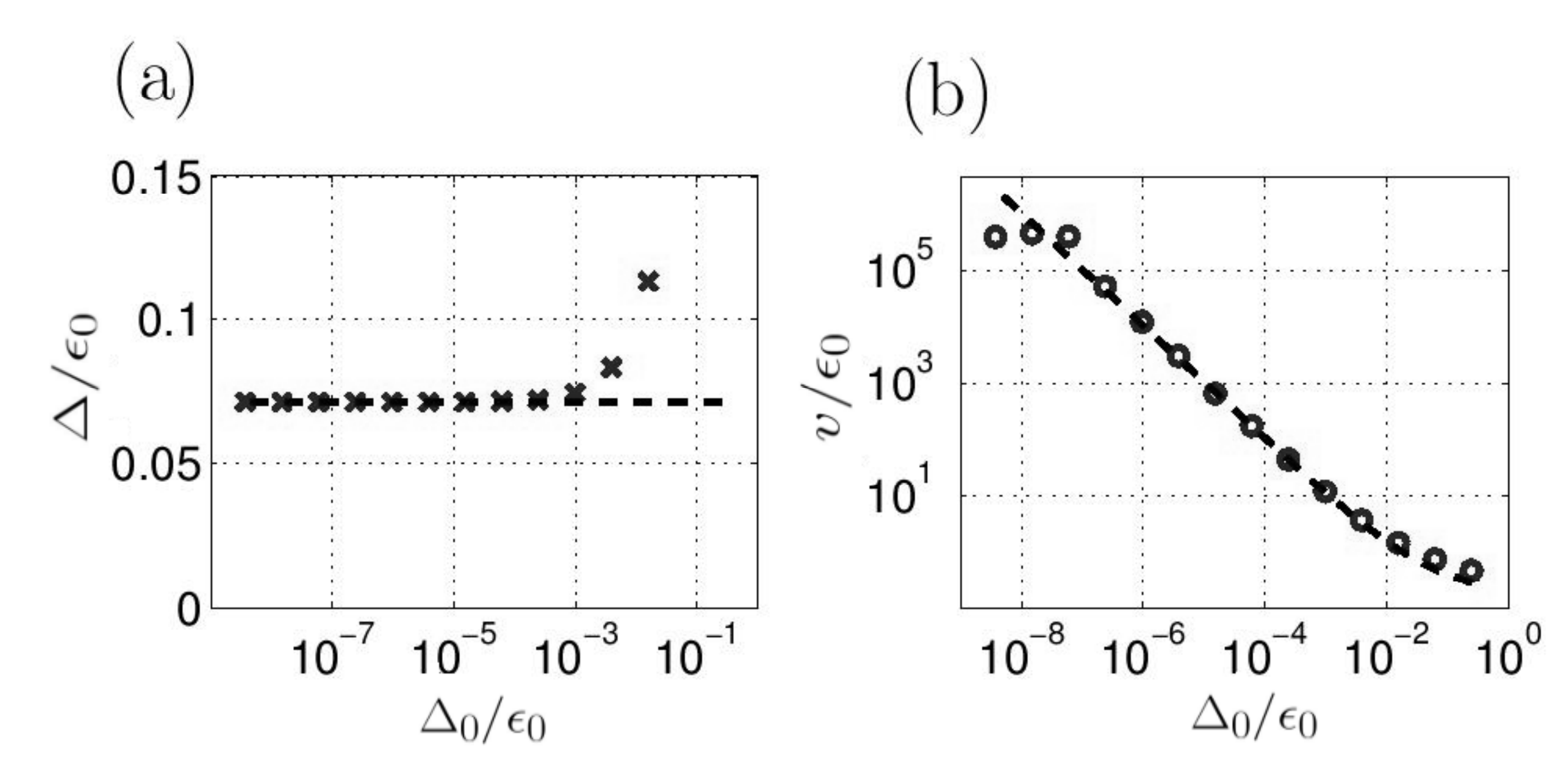}
\caption{External field dependence of (a) order parameter and (b) couplings for the rBCS model \cite{lauscher}.} 
\end{figure}

Integrating these flow equations one arrives at a value of the final gap that, obviously, depends on the value of the external field, as shown in Fig. \ref{rbcsresults} on the left panel. In order to obtain good results it is thus desirable to use as small fields as possible. Unfortunately, there is a trade-off, shown on the right panel of Fig. \ref{rbcsresults}: As one lowers the value of the external field, the couplings grow larger. In this case, the growth of the couplings represents the massless Goldstone mode associated to phase fluctuations of the order parameter\cite{lauscher}. 

 As the flow equations for this model are exact, large couplings (i.e. four point vertices) do not really matter. For more realistic problems, however, it is crucial to keep the couplings at moderate values because the truncation of the flow equations corresponds to a weak coupling approximation: Terms of order $V^3$, where $V$ are the couplings, are neglected. Thus the use of external fields puts one into the dilemma of either getting errors due to the unphysical field or due to large couplings. In order to meliorate the situation we will try a slightly different approach with counterterms.

\section{Interaction Flow and Counterterms} \label{ch:counterterms}

In this section we discuss the introduction of counterterms into the fRG scheme as suggested by Gersch et al.\cite{gersch}. In order to do this it is necessary to abandon sharp momentum cutoffs (as will be explained below) and to use a soft or even flat cutoff instead. We will always use the so-called \emph{interaction flow} \cite{intflow}, so this will be topic of the next section before actually introducing the counterterm method.

\subsection{Interaction Flow} \label{sec:intflow}
 \noindent The interaction (or flat-cutoff) flow scheme is implemented by introducing the following flow parameter dependence of the quadratic part of the action, $Q$,
\begin{equation}
Q\rightarrow Q_\chi = \frac{Q}{\chi},
\end{equation}
where $\chi$ is a $k$-independent real number $\in [0,1]$. $\chi = 0$ is the starting point of the flow. As the Green's function of the virtual fluctuations is $Q_\chi^{-1}$, this cutoff choice increases the weight of the fluctuations continuously from 0 to 1 for all $k$, i.e. corresponds to a completely flat cutoff.
Recalling the definition of the full action,
\begin{equation}
S_\chi[\Psi] = \left(\Psi,Q_\chi \Psi\right) + V[\Psi],
\end{equation}
and assuming that $V[\Psi]$ is quartic in the fields, one finds by rescaling
\begin{equation}
\Psi = \sqrt{\chi}\tilde{\Psi}
\end{equation}
that this choice of flow parameter amounts to switching on the interactions gradually,
\begin{equation}
\tilde{S}_\chi[\tilde{\Psi}] = \left(\tilde{\Psi},Q \tilde{\Psi}\right) + \chi^2 V[\tilde{\Psi}].
\end{equation}
Thus the interaction flow scheme meets the criteria for a sensible flow scheme as (a) the partition function at $\chi=0$ can be evaluated exactly and (b) $\chi=1$ corresponds to the physical system of interest. The reason for calling this scheme a cutoff scheme with an infinitely soft cutoff is that $\chi$ is a constant function in momentum space and may hence be viewed as a cutoff function in momentum space in the limit where the slope of the cutoff is sent to zero \cite{intflow}.

\subsection{Counterterms}
 
\noindent The idea of introducing counterterms is simply to add zero in the form
\begin{equation}
0 = a - a
\end{equation}
to the Hamiltonian (or, equivalently,to the action) of a quantum system. This is to obtain a different separation of the Hamiltonian into a 'free' and an 'interacting' part, explicitly,
\begin{eqnarray}
\mathcal{H} = \mathcal{H}_0 + \mathcal{H}_I \\ \nonumber
\mathcal{H}_0 \rightarrow \mathcal{H}_0 + \delta\mathcal{H} \\ \nonumber
\mathcal{H}_I \rightarrow \mathcal{H}_I - \delta\mathcal{H}.
\label{dH}
\end{eqnarray}
For this operation to to be useful, of course the counterterm $\delta\mathcal{H}$ should be quadratic in the fields. This change is very similar to choosing a different expansion point for perturbation theory\cite{neumayr}. 

Following Gersch et al. \cite{gersch}, we introduce counterterms in order to break a symmetry explicitly during the flow as in the method using small external fields described above. The difference to the introduction of an external field is that with the counterterm the symmetry of the full Hamiltonian is preserved, so that no error is introduced into the calculation of the corresponding order parameter. hence the counterterm does not need to be small.
This last point, however, only holds if the generating functional of the problem is evaluated \emph{exactly}, because if and only if this is the case the two copies of $\delta\mathcal{H}$ introduced in Eq. \ref{dH} cancel. 

To see how this works in practice, the rBCS model already introduced in section \ref{sec:rbcs} will serve as a simple example. It has the additional advantage that the truncated flow equations for this model are \emph{exact}, so that the result will not depend on the choice of counterterm (more precisely, its numerical value). In fact, the flow equations for the cutoff flow with an external field, eq.(\ref{rbcsflow}) can be taken over almost literally. To recall them, these equations are given by
\begin{eqnarray}
\dot{v} &=& \dot{B}\left(v^2 + w^2\right) + 2 \dot{A}vw \\
\dot{w} &=& \dot{A}\left(v^2 + w^2\right) + 2\dot{B}vw\\
\dot{\Sigma} &=& C\left(v+w\right) \label{rbcsflow}
\end{eqnarray}
where $\dot{A},\dot{B}$ and $C$ refer to the loop integrals given by
\begin{eqnarray}
A &=& -T\sum_\omega\int dk \left(\frac{\Sigma\chi(s)^2}{|Q|^2 + (\Sigma\chi(s))^2}\right)^2 \\
B &=& T\sum_\omega\int dk \frac{(\omega^2 + \epsilon_k^2)\chi(s)^2}{\left(|Q|^2 + (\Sigma\chi(s))^2\right)^2}\\
C &=& T\sum_\omega\int dk \frac{2\chi(s)\dot{\chi(s)}\Sigma|Q|^2}{|Q|^2 + (\Sigma\chi(s))^2}.\label{rbcsloops}
\end{eqnarray}
Here $s$ is the flow parameter, whereas $\chi(s)$ is a cutoff function in momentum space. 
Now we use the $s$-dependence of $\chi(s)$ corresponding to the interaction flow\begin{equation}
\chi(s) = s, \qquad s\in [0,1].
\end{equation}
\begin{figure}[h]
\centering
\includegraphics[width = 10cm]{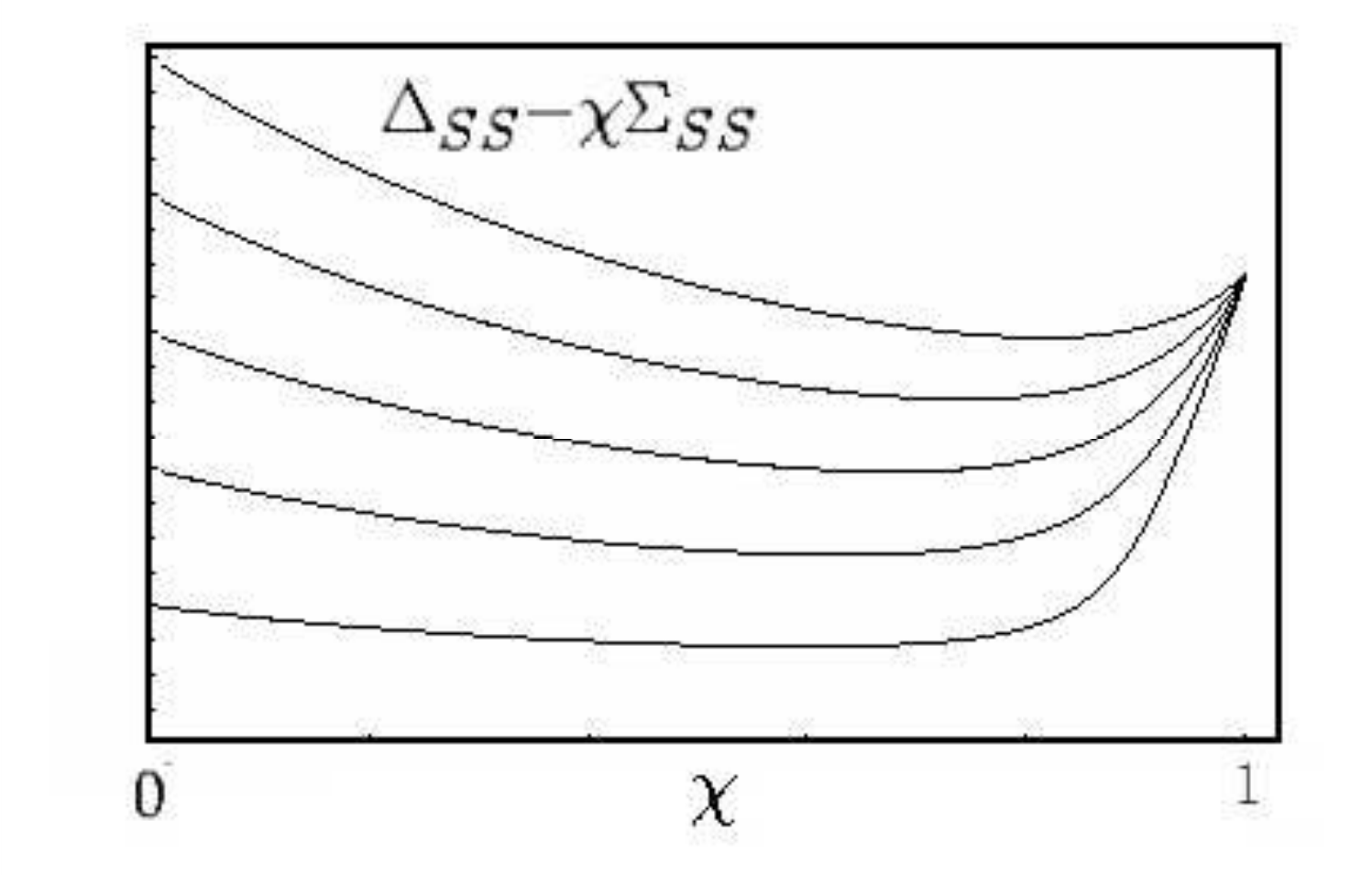}
\label{rbcs}
\caption{Flow of the effective gap for the rBCS model. The gap flows to the same value for all choices of counterterms.}
\end{figure}
Next, in order to cancel the external field (or initial self-energy) $\Sigma(\chi =0)$ from the action, the counterterm $\Delta$ is added to the quadratic part $Q$, i.e. $Q \to Q + \Delta$. Remembering that
\begin{equation}
\mathcal{G}^{-1} = Q - \Sigma,
\end{equation}
it is clear that this term $\Delta$ must be given by $\Delta = \Sigma (\chi=0)$. Thus, in order to get the flow equations for the symmetric situation with counterterms from the equations with symmetry-breaking externel fields, one only has to augment the free inverse propagator, $Q$, by an anomalous term $\Delta$, and the self-energy in the full propagators, which was just $\chi \Sigma$ before, has to be shifted according to 
\begin{equation}
\chi\Sigma \rightarrow \chi\Sigma - \Delta,
\end{equation}
The expression on the right hand side is called the \emph{effective gap}. These changes do not change eqns.(\ref{rbcsflow}), but eqns.(\ref{rbcsloops}) now read
\begin{eqnarray}
A &=& -T\sum_\omega\int dk \left(\frac{(\Sigma\chi-\Delta)\chi}{|Q|^2 + (\Sigma\chi-\Delta)^2}\right)^2 \\
B &=& T\sum_\omega\int dk \frac{(\omega^2 + \epsilon_k^2)\chi(s)^2}{\left(|Q|^2 + (\Sigma\chi(s)-\Delta)^2\right)^2}\\
C &=& T\sum_\omega\int dk -\frac{\Delta^3 - 2\chi\Sigma(\Delta^2+\omega^2+\epsilon_k^2) + \Delta\left((\chi\Sigma)^2 + \omega^2 + \epsilon_k^2\right)}{|Q|^2 + (\Sigma\chi(s)-\Delta)^2}.
\label{rbcsloopsc}
\end{eqnarray}
These flow equations may now be integrated, which in fact leads to a result for the gap that is \emph{independent} of the choice of counterterm, and which is not affected by the application of an unphysical external field. In Fig.  \ref{rbcs} we show some flows for different values of the counterterm.

We note that an alternative way to express this recipe is the following: First one adds the counterterm $\Delta$ to the quadratic part of the self-energy. Then one sets the initial condition $\Sigma (\chi=0)$ equal to $\Delta$, such that in the full inverse propagator for $\chi=1$ the effective gap $\chi \Sigma - \Delta$ is exactly cancelled. However in the flow before reaching $\chi=1$, the systems appears to be in the symmetry-broken state. This allows to uncover the spontaneous symmetry breaking. 

For the reduced BCS model, the counterterm methods works just as well as the external field scheme of Ref. \onlinecite{lauscher}. In many cases\cite{gersch}, the counterterm method has the apparent advantage that large couplings can be avoided for a large part of the flow, or more precisely, by keeping the flow in the symmetry-broken state for most of the flow range, criticality can be avoided. This works best for systems without Goldstone bosons (e.g. discrete symmetry breakings). But even for breaking of continuous symmetries where the Goldstone modes require a diverging component of the vertex at the end of the flow, avoiding the criticality of the amplitude modes that are usually gapped at the end of the flow could represent an improvement.
Of course, the valid question is whether also non-trivial models can be treated correctly with this method. In the following we will apply the counterterm flow to a model in one spatial dimension with competing ordering tendencies.

\section{Attractive onedimensional model - Bosonization treatment} \label{sec:tomonaga}
\subsection{Model}
\noindent In general, the action for free spin-$1/2$ fermions in one dimension at $T = 0$ reads
\begin{equation}
S_F = \sum_{s=\uparrow, \downarrow} \int \difd x \int_0^\beta \difd \tau \, \bar{\psi}_s(x,\tau) (\partial_\tau + \epsilon_k - \mu)\psi_s (x,\tau),
\end{equation}
where $\epsilon_k$ is understood as a differential operator defined by the identification $k \rightarrow i\partial_x$. For low energies, $E \ll \mu$, the dispersion may be linearized around the Fermi points, yielding
\begin{equation}
\epsilon_{k_F + q} \approx \epsilon_{k_F} + \partial_k \epsilon_k\Big|_{k=k_F} q \equiv \mu + v_F q,
\end{equation}
where $v_F$ is the Fermi velocity defined by above relation. Note also that $v_F$ at the left Fermi point differs in sign from $v_F$ at the right Fermi point. In order to write the effective low-energy action it is convenient to define chiral components of $\psi$ as
\begin{eqnarray}
R_s(x,\tau) = \sum_{k > 0} e^{i k x} \psi_k(\tau) \\
L_s(x,\tau) = \sum_{k < 0} e^{i k x} \psi_k(\tau) \nonumber
\end{eqnarray}
where $\psi_k(\tau)$ denotes the Fourier transform of $\psi(x,\tau)$ with respect to $x$. These new fields describe right and left moving electrons, respectively, and in terms of them, the action can be written as
\begin{equation}
S_F = \sum_{s=\uparrow, \downarrow} \int \difd x \difd \tau R_s^\dagger(x,\tau)(\partial_\tau + i v_F \partial_x)R_s(x,\tau) + L_s^\dagger(x,\tau)(\partial_\tau - i v_F \partial_x) L_s(x,\tau) 
\end{equation}
where now $v_F$ is taken to be positive and inversion symmetry is assumed. Crossterms between $R_s(x,\tau)$ and $L_s(x,\tau)$ are neglected.

\begin{figure}[tb]
\centering
\includegraphics[width = 13cm]{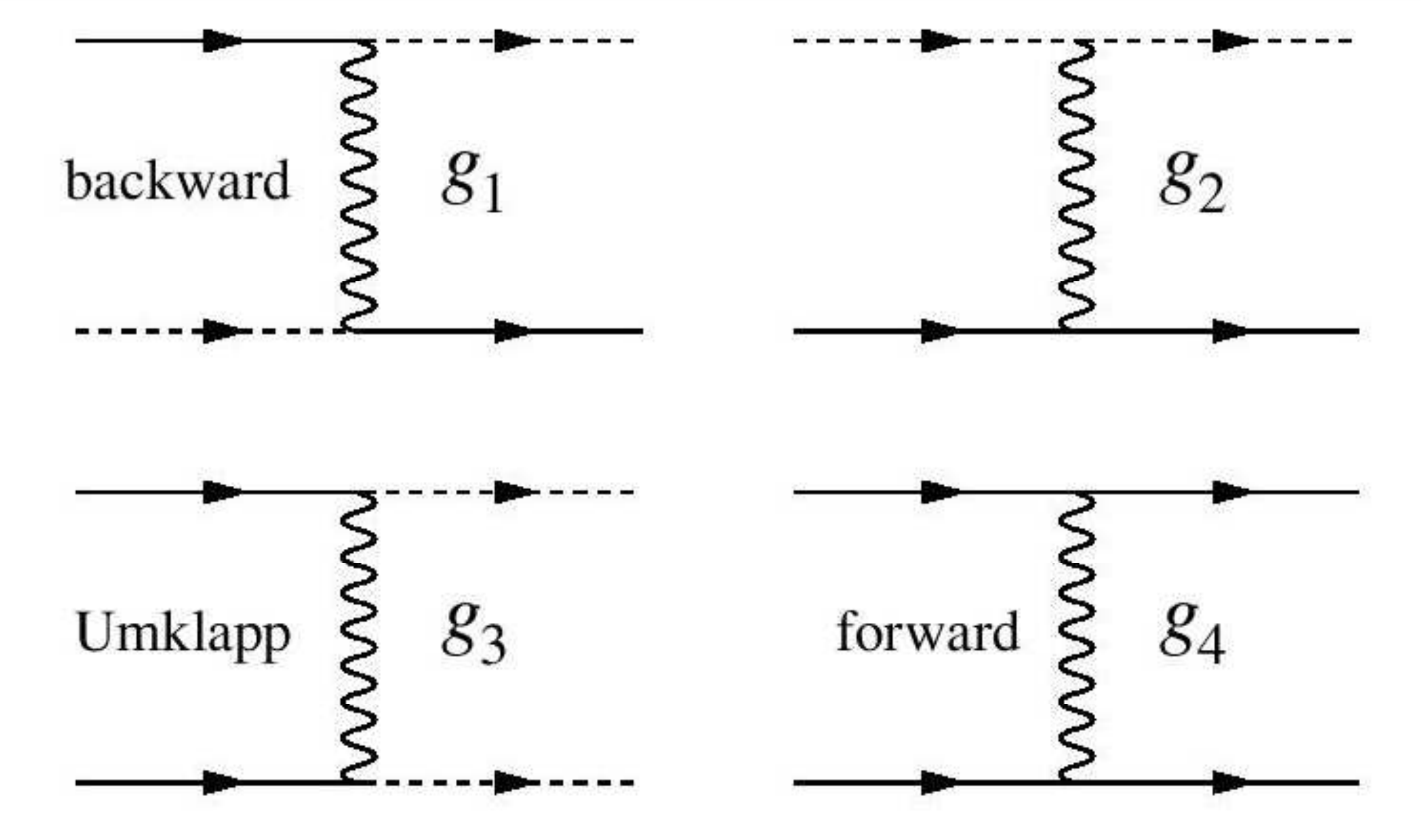}
\caption{Classification of effective low energy interaction vertices in the Tomonaga-Luttinger model. Solid lines are located in the vicinity of one Fermi point (e.g. right), dashed lines in the vicinity of the other (e.g. left) \cite{senechal}.}
\label{interactions}
\end{figure}

Now we include interactions. As we are looking for a situation with competing orders, the attractive Tomonaga-Luttinger model at \emph{half-filling} will be in the focus of interest. Half band filling implies that at arbitrarily small energies umklapp scattering is an allowed process and will be consequently included in the calculations. Traditionally, the interactions for this situation are classified according to the \emph{g-ology}-scheme into various channels, see Fig. \ref{interactions}. In the following, the $g_4$ processes are neglected, because they are not relevant for symmetry breaking. Furthermore, the interaction vertices in Fig. \ref{interactions} do not carry spin-labels for the in- and outgoing legs. For spin-$1/2$ fermions, each vertex can in principle occur with parallel or perpendicular spins for the ingoing (outgoing) legs. However, because of the Pauli principle the process with parallel spins is only allowed if other quantum numbers for the fermions involved differ. This means that for the $g_3$ process only the perpendicular process is allowed as both ingoing (outgoing) legs are located at the same Fermi point. Also  $g_{1\parallel} = -g_{2\parallel}$ holds in general so that these two vertices actually correspond to the same vertex. This leads to the set of vertices $\{g_1,g_2,g_\parallel,g_3\}$, where the $g_i,\,( i = 1,2,3)$ refer to the perpendicular process in the corresponding channel. If the couplings are invariant under spin rotations, $g_\parallel = g_2-g_1$. The interaction Hamiltonian with this notation (and without $g_4$) is given by
\begin{eqnarray*}
\mathcal{H}_I &=& v_F\int\mathrm{d} x \left[  g_1\sum_\sigma L^\dagger_\sigma R^\dagger_{-\sigma} L_{-\sigma} R_\sigma - g_2\sum_\sigma L^\dagger_\sigma R^\dagger_{-\sigma} L_\sigma R_{-\sigma} \right.  \\ & & \left. - g_{\parallel}\sum_\sigma L^\dagger_\sigma R^\dagger_\sigma L_\sigma R_\sigma + \frac{g_3}{2}\sum_\sigma \left ( L^\dagger_\sigma L^\dagger_{-\sigma} R_{-\sigma} R_{\sigma} + R^\dagger_\sigma R^\dagger_{-\sigma} L_{-\sigma} L_{\sigma}\right) \right]. \label{interaction}
\end{eqnarray*}

\subsection{Bosonization}
In the bosonization procedure\cite{giamarchi,senechal}, fermionic operators for left- and right-movers are expressed by exponentials of a bosonic field $\phi$. Omitting the spin index for a moment, one writes
\begin{eqnarray}
R,\bar{R} \qquad&\longleftrightarrow&\qquad \frac{1}{\sqrt{2\pi a}} \exp\left[\pm i\sqrt{4\pi}\phi(x,\tau)\right], \nonumber \\
L,\bar{L} \qquad&\longleftrightarrow&\qquad \frac{1}{\sqrt{2\pi a}} \exp\left[\mp i\sqrt{4\pi}\bar{\phi}(\bar{x,\tau})\right],
\label{bosotable}
\end{eqnarray}
For systems with spin-1/2 one needs two bosons $\phi_\alpha$, $\alpha \in \{\uparrow,\downarrow\}$, instead of one. However, it turns out to be more convenient to define instead
\begin{eqnarray}
\phi_\sigma = \phi_\uparrow - \phi_\downarrow, \\
\phi_\rho = \phi_\uparrow + \phi_\downarrow.
\end{eqnarray}

Using the bosonization formulae, Eq. \ref{bosotable}, the bosonized form of the full interacting Hamiltonian is readily found to be
\begin{equation}
\ham = \ham_0 + \ham_I = \ham_\sigma + \ham_\rho. \label{bosham}
\end{equation}
Here $\ham_\rho(\ham_\sigma)$ depends on $\phi_\rho(\phi_\sigma)$ only, which means that these two fields actually decouple. The expressions for the $\ham_\mu,\quad \mu = \rho,\sigma$ are given by
\begin{eqnarray}
\ham_\rho &=& \frac{u_\rho}{2}\int\mathrm{d}x\left [  K_\rho \Pi^2_\rho + \frac{1}{K_\rho}(\partial_x\phi_\rho)^2\right ] + \frac{u_\rho g_3}{2\pi^2}\cos\left(\sqrt{8\pi}\phi_\rho\right ), \\
\ham_\sigma &=& \frac{u_\sigma}{2}\int\mathrm{d}x\left [  K_\sigma \Pi^2_\sigma + \frac{1}{K_\sigma}(\partial_x\phi_\sigma)^2\right ] + \frac{u_\sigma g_1}{2\pi^2}\cos\left(\sqrt{8\pi}\phi_\sigma\right ),
\end{eqnarray}
and the constants $K_\mu$ and $u_\mu$ are
\begin{equation}
K_\mu = \sqrt{\frac{2\pi - (g_\parallel \mp g_{2})}{2\pi + (g_\parallel \mp g_{2})}}\qquad u_\mu = v_F\sqrt{1-\left(\frac{g_{2} \mp g_\parallel}{\pi}\right )^2},
\end{equation}
where the upper(lower) sign refers to $\mu = \sigma$($\mu = \rho$). In particular we note for later use that for $g_1<0$, $K_\sigma < 1$, while $K_\rho$ can be tuned to be larger than 1 by making $g_\parallel+g_2$ negative. 

The $\ham_\mu$ have the same structure, they are both of the well-known sine-Gordon form, which implies in particular that they exhibit a phase-transition of the Kosterlitz-Thouless type as the parameters (i.e. the couplings) are varied \cite{giamarchi}. This phase transition can be analysed using the Renormalization Group, leading to the phase diagram for the Tomonaga-Luttinger model. This will be the content of the next subsection.

\subsection{Phase Diagram} \label{sec:phasediagram}
\begin{figure}[tb]
\centering
\includegraphics[width = 13cm]{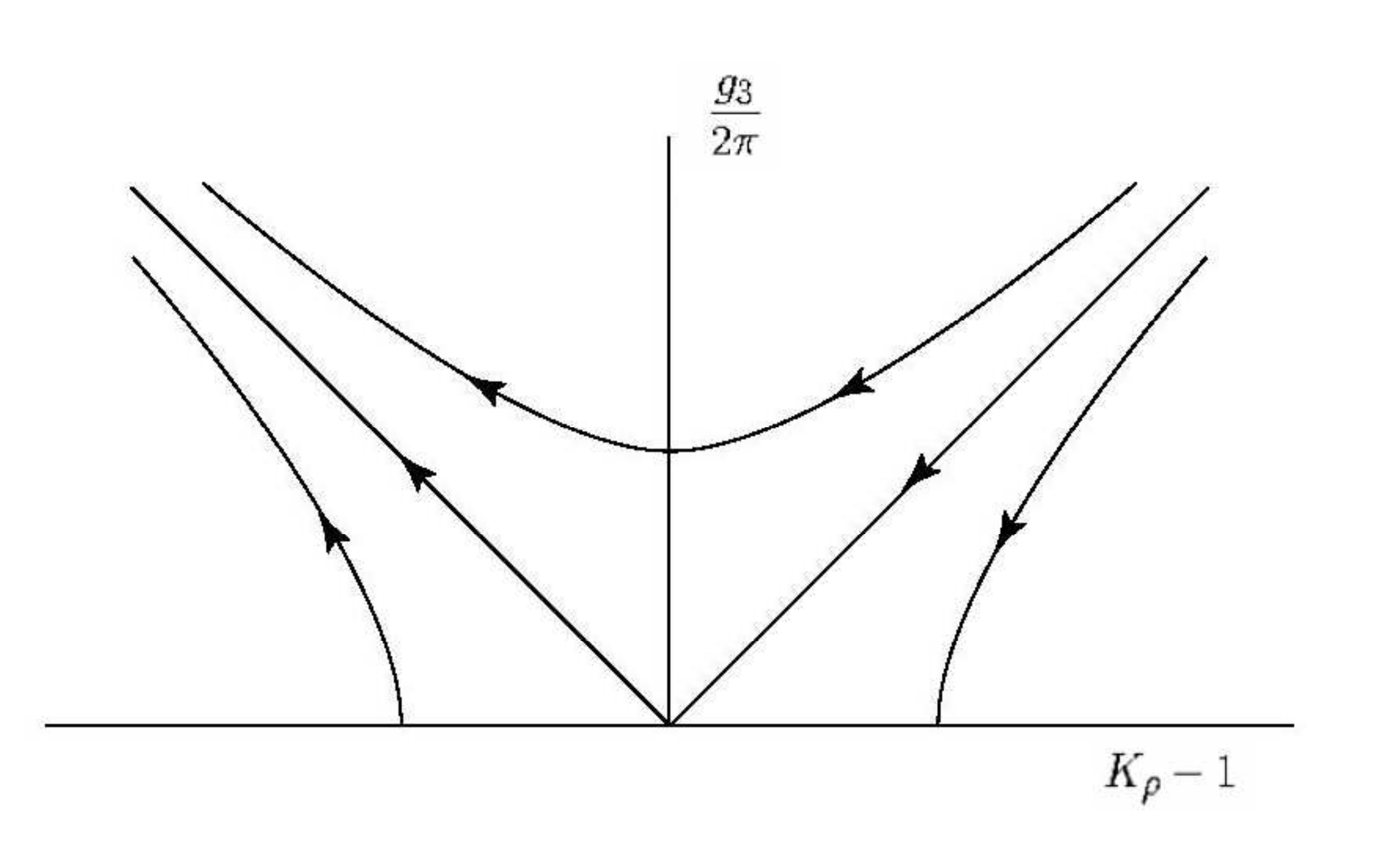}
\label{rgflow}
\caption{RG flow in the charge sector near $K_\rho=1$. For a massive spin sector, the diagonal in the right half is the separation line between the phases with leading $s$SC correlations ($g_3$ flows to zero) and CDW order in the ground state ($g_3$ is relevant). }
\end{figure}

The derivation of the one-loop RG equations for the sine-Gordon models of Eq. \ref{bosham} can be found for example in Ref. \onlinecite{giamarchi}. As spin and charge sector decouple, there are two independent sets of flow equations, which in the charge sector are given by 
\begin{equation}
\frac{\difd K_\rho}{\difd l} = -\frac{1}{2\pi^2}K_\rho^2g_3^2\qquad \frac{\difd g_3}{\difd l} = - 2g_3(K_\rho-1). \label{krhogflow}
\end{equation}
The equations for the spin sector are obtained by replacing $\rho$ by $\sigma$ and $g_3$ by $g_1$ everywhere. The flow generated by these equations can easily be seen to lead to a phase transition by plotting it in the $g_3-K_\rho$ plane, as shown in Fig. \ref{rgflow}. The flow for negative $g_3$ is analoguous, as can be inferred from Eqs. \ref{krhogflow}.
Whilst the cosine term appears to be irrelevant for $K_\rho<1$ on dimensional grounds, it may actually be marginally relevant because $K_\rho$ is renormalized during the flow. This leads to the separatrix given by the straight lines in Fig. \ref{rgflow}, separating a phase where $g_3$ flows to strong coupling from a phase where it flows to zero. The system flows to strong coupling if
\begin{equation}
|g_3| > 2\pi(K_\rho-1).
\end{equation}
In the bosonization picture, the flow to strong coupling can be interpreted physically as the development of a gap in the charge boson spectrum. This follows from the fact that as $g_3$ becomes large, the cosine potential becomes steeper and steeper, so that the field $\phi_\rho$ will get trapped in one of the minima, because the kinetic energy is no longer high enough to overcome the energy barriers separating the minima. Of course, one could object that tunneling between the minima via instantons might spoil this picture. However, it can be shown that finite energy instantons for the 1D sine-Gordon model do not exist \cite{rajaraman}. For large $g_3$, the cosine may be expanded around the minimum, leading to a mass term (i.e. a gap) for $\phi_\rho$,
\begin{equation}
\frac{v_F g_3}{\pi^2}\cos\left(\sqrt{8\pi }\phi_\rho\right) \approx \mbox{const.} + \frac{8  v_F g_3}{\pi}\phi_\rho^2 \equiv \mbox{const.} + m\phi_\rho^2,
\end{equation}
where the mass, $m$, is defined by the above relation.

Having found the low-energy behavior in the bosonic picture, the correlations in the fermionic picture can be discussed. Generically, one or more order parameters develop power law correlations. Due to the Mermin-Wagner theorem in one dimension continuous symmetries can not be broken (even at $T=0$), so that no long range order develops if no discrete symmetry breaking is involved. 
Instead, one finds a scale-free state, a Luttinger liquid, that may be viewed as a system on the verge of various symmetry breaking instabilities. Typically, the degree of divergence of the static susceptibilities is not the same for the different instabilities, so that one instability is dominant. When coupled to a higher dimensional environment, the system would order according to its dominant instability \cite{giamarchi}. The susceptibilities can be expressed in terms of correlations of the corresponding order parameters, which in turn can be given in terms of the bosonic fields. Thus we will start with the bosonized expressions of the two order parameters that will be of interest later: $s$-wave superconductivity ($s$SC) and charge density wave (CDW).

From the bosonization rules introduced above, it is easy to verify that the order parameters for CDW and $s$SC are given by
\begin{eqnarray}
\ocdw(x) = \frac{e^{-i k_F x}}{\pi a} e^{i\opf\phi_\rho}\cos\left(\opf\phi_\sigma\right), \\
\oss(x) = \frac{1}{\pi a} e^{-i\opf\theta_\rho}\cos\left(\opf \phi_\sigma\right).
\end{eqnarray}
The correlation functions for these order parameters can be obtained from the correlators of the various vertex operators appearing in their definition. They read
\begin{eqnarray}
\langle \ocdw^\dagger(r)\ocdw(0)\rangle = \frac{e^{2i k_F x}}{2(\pi a)^2} \left(\frac{a}{r}\right)^{K_\rho + K_\sigma}, \nonumber \\
\langle \oss^\dagger(r)\oss(0)\rangle = \frac{1}{(\pi a)^2} \left(\frac{a}{r}\right)^{K_\rho^{-1} + K_\sigma}.
\label{ocorrelations}
\end{eqnarray}
The behavior of the susceptibilities can be directly calculated from the order parameter correlations, leading to the singular behavior
\begin{eqnarray}
\chi_{\mbox{CDW}}(k=0,\omega) \propto \omega^{K_\rho + K_\sigma - 2} \\
\chi_{\mbox{$s$SC}}(k=0,\omega)\propto \omega^{K_\rho^{-1} + K_\sigma - 2}.
\end{eqnarray}
It should be noted that above formulae hold if both $\phi_\sigma$ and $\phi_\rho$ are massless. Of course, $K_\rho$ and $K_\sigma$ are renormalized, so actually one should first perform the RG calculation and afterwards use the renormalized values to calculate the singular behaviour of the susceptibilities. 

The field $\phi_\sigma$ will be massive if $|g_1| > 2\pi (K_\sigma -1)$ which is certainly the case for $K_\sigma<1$, i.e. $g_\parallel - g_2 > 0$. Later, in the fRG treatment, we will focus on this regime.
Then the behavior of the spin-sector is modified as compared to the massless case. A massive $\phi_\sigma$ implies that operators like $\cos \phi_\sigma$ acquire nonzero expectation values, or that their correlators are strongly suppressed, depending on their value at the potential minimum. For the attractive case, $g_1 < 0$, it turns out that all order parameter correlations apart from $s$SC and CDW are suppressed if $\phi_\sigma$ orders \cite{giamarchi}. The correlations of $\ocdw$ and $\oss$ in this case are modified according to
\begin{eqnarray}
\langle\ocdw^\dagger(r)\ocdw(0)\rangle \propto \left(\frac{a}{r}\right)^{K_\rho} \nonumber\\
\langle\oss^\dagger(r)\oss(0)\rangle \propto \left(\frac{a}{r}\right)^{K_\rho^{-1}},
\end{eqnarray}
and the same modification occurs for the static susceptibilities,
\begin{eqnarray}
\chi_{\mbox{CDW}} \propto \omega^{K_\rho-2} \nonumber \\
\chi_{\mbox{$s$SC}} \propto \omega^{K_\rho^{-1}-2}.
\end{eqnarray}
This obviously originates in the fact that a massive field can no longer fluctuate at arbitrarily low energies, so that $\phi_\sigma$ can no longer influence the asymptotic behavior of the system. 

$s$SC will be the dominant instability as long as $\phi_\rho$ remains massless, because in this regime $K_\rho \geq 1$ always holds (where equality can only occur for $g_3=0$), so that $\chi_{\mbox{$s$SC}}$ always diverges stronger than $\chi_{\mbox{CDW}}$.

What happens if $\phi_\rho$ becomes massive, too? The answer can be read off the bosonized form of the order parameters. $\oss$ depends on the dual field, $\theta_\rho$. If $\phi_\rho$ orders, its dual field has to disorder, so that its correlations are exponentially suppressed \cite{giamarchi}, and the $s$SC susceptibility is no longer singular. $\ocdw$, on the other hand, acquires a finite value as it  $\phi_\rho$. At half-filling, CDW only breaks $\mathbb{Z}_2$ which is a discrete group, so there are no low-lying modes that could destroy long-range order, so that at $T=0$, CDW ordering occurs.

To summarize, bosonization for our half-filled attractive model with massive spin sector predicts that the transition from massless to massive $\phi_\rho$, tuned by crossing the separation line in Fig. \ref{rgflow} either by increasing $|g_3|$ or decreasing $K_\rho$, is accompanied by a transition from a phase with dominant $s$SC instability \emph{and} subdominant  CDW power-law correlations to a phase with CDW ordering and exponentially decaying $s$SC correlations.

\section{Functional Renormalization Group Approach}

In this section, the fRG with counterterms will be applied to the attractive Tomonaga-Luttinger model with interactions given by Eq. \ref{interactions} at half-filling. We focus on the attractive case with massive spin sector, as this gives us the possibility to study the competition between charge-density-wave and superconducting ordering tendencies. The goal is  to test the applicability of the fRG with symmetry-breaking counterterms to systems  where there are competing correlations, and where the truncated flow equations are not exact. 
The bosonization phase diagram established in the last section will be the guideline for checking the quality of the results obtained. 
A word of caution is appropriate in the beginning: As shown above, the different phases of 1D systems do not correspond to different ordered states in general, but to critical states with one or several diverging static susceptibilities. In the currently used setup, the fRG  \emph{cannot} reproduce this critical behaviour. Instead, because order parameter fluctuations are not included, one arrives at the ordered state corresponding to the strongest instability. For this reason, numerical values like the order parameter should be understood as energy scales, and as indicator for the dominant correlation. The parameters found by fRG  for the transition point from the superconductivity dominated phase to the charge density wave phase are used as a check for the correctness of the results. 
We see the main strength of the method tested here in two-dimensional applications, where long-range order in the ground state is more frequent than in one dimension. Hence the mischaracterization of  power-law odered states as long-range orderd states is for us a tolerable defect which is clearly rooted in the nature of the approximation.

\subsection{Counterterms and Propagator}
As the model exhibits two different instabilities, we will need two counterterms to deal with them and to render the flows finite in the whole parameter range. These are the anomalous $s$SC self-energy already discussed in Sec. \ref{sec:rbcs} and the counterterm corresponding to a commensurate charge density wave (CDW). The order parameter for CDW symmetry breaking at half-filling, $\Delta_{\mbox{CDW}}$ couples to the fermions via
\begin{equation}
H_{\mbox{CDW}}= \Delta_{\mbox{CDW}} \sum_\sigma\sum_k c^\dagger_{k,\sigma}c_{k+Q,\sigma},
\end{equation}
where $Q = 2 k_F = \pi$ for half filling . In the language of chiral fermions, the relations
\begin{equation}
R_{k\pm Q} = L_{k}, \qquad L_{k\pm Q} = R_{k}
\end{equation}
hold, so that $H_{\mbox{CDW}}$ can be written as
\begin{equation}
H_{\mbox{CDW}}= \Delta_{\mbox{CDW}}  \sum_\sigma\sum_k \left\{R^\dagger_{k,\sigma}L_{k,\sigma} + L^\dagger_{k,\sigma}R_{k,\sigma}\right\}.
\end{equation}
Using similar relations, one finds for the pairing part $H_{\mbox{$s$SC}}$ that
\begin{equation}
H_{\mbox{$s$SC}} = \Delta_{\mbox{$s$SC}}  \sum_k \left\{L^\dagger_{k,\uparrow}R^\dagger_{-k,\downarrow} + R^\dagger_{k,\uparrow}L^\dagger_{-k,\downarrow} + \mbox{h.c.}\right\}.
\end{equation}
Next, we introduce a Nambu spinor notation for the fields, so that the free inverse propagator can be summarized in a $4\times 4$ matrix structure. To this end, one must take care to arrange the fields in the spinors in a way such that $\bar{\Psi}$ and $\Psi$ are independent from each other, because otherwise the mathematical structure of the action is changed \cite{nakahara}. The choice of spinors in this work is
\begin{equation}
\Psi_k = \left(\begin{array}{c} R_+ \\ \bar{L}_- \\ L_+ \\ \bar{R}_-\end{array}\right),\qquad \bar{\Psi}_k = \left(\bar{R}_+, L_-, \bar{L}_+, R_-\right),
\end{equation}
where $+$ is shorthand for $(k,\uparrow)$ and $-$ means $(-k,\downarrow)$. The inverse free propagator including counterterms reads with these conventions
\begin{equation}
Q = \left(\begin{array}{cccc}Q^R_k & \Delta_{sSC} & \Delta_{CDW}& 0 \\ \Delta_{sSC} & \tilde{Q}^L_{-k} & 0 & -\Delta_{CDW} \\ \Delta_{CDW} & 0 & Q^L_k & \Delta_{sSC} \\ 0 & -\Delta_{CDW} & \Delta_{sSC} & \tilde{Q}^R_{-k}\end{array}\right ), 
\end{equation}
where 
\begin{equation}
Q_k^R = Q^L_{-k} = i\omega - v_F k,\qquad \tilde{Q}^R_k = \tilde{Q}^L_{-k} = i\omega + v_F k
\end{equation}
represent the symmetric inverse free propagator, and $\tilde{Q}$ is the transpose of $Q$. 

Introducing the flow parameter $\chi \in [0, 1]$ for the interaction flow (as in Subsection \ref{sec:intflow}), the full propagator is given by the standard formula $\mathcal{G} = \left(\frac{1}{\chi}(Q - \chi\Sigma)\right)^{-1}$. Here $\Sigma$ is again reserved for the flowing selfenergy. Herein we take the counterterms $\Delta_{CDW}$ and $\Delta_{sSC}$ as initial conditions at $\chi=0$. This exactly cancels the $\Delta$s added to $Q$ for $\chi =1$, so that in the end the model is unchanged.
This leads to the expression
\begin{eqnarray}
\mathcal{G} &=& \frac{\chi}{\omega^2 + (\Delta_{sSC}-\chi\Sigma_{sSC})^2 + (\Delta_{CDW}-\chi\Sigma_{CDW})^2}\times \\ & &\nonumber\left (\begin{array}{cccc} -i\omega + -\epsilon_k  & \Delta_{sSC}-\chi\Sigma_{sSC} & \Delta_{CDW} - \chi\Sigma_{CDW} & 0 \\  \Delta_{sSC}-\chi\Sigma_{sSC} & -i\omega + \epsilon_k & 0 & -\Delta_{CDW}+\chi\Sigma_{CDW} \\ \Delta_{CDW}-\chi\Sigma_{CDW} & 0 & -i\omega + \epsilon_k & \Delta_{sSC}-\chi\Sigma_{sSC} \\ 0 & -\Delta_{CDW}+\chi\Sigma_{CDW} & \Delta_{sSC}-\chi\Sigma_{sSC} & -i\omega -\epsilon_k\end{array}\right ).
\end{eqnarray}
for the full propagator that occurs in the flow equations. One should note that the elements of the counter diagonal are all identically zero, implying that, apart from the symmetry breaking propagator entries for $s$SC and CDW, no new anomalous entries are created. The new propagator stemming from the CDW symmetry breaking correlates left- and right-movers of the same spin and will thus subsequently be called umklapp propagator. 

 The single-scale propagator is obtained from the definitions of $\mathcal{G}$ and $Q$ above from its definition (cf. eq.(\ref{defsprop}),
\begin{displaymath}
S = -\mathcal{G}\dot{Q}\mathcal{G},
\end{displaymath}
where the product here is of course given by matrix multiplication. The entries with indices $ij$ for which 
\begin{displaymath}
Q_{ij} = \Delta_{\mbox{X}}
\end{displaymath}
holds will be called $S_{\mbox{X}}$ with X = $s$SC, CDW.

 Of course, as already mentioned in section \ref{sec:rbcs}, the off-diagonal entries of the propagator may generate new vertices during the flow. In the case at hand, however, no new vertices are generated apart from the anomalous superconducting vertex introduced in section \ref{sec:rbcs}. This vertex will be named $g_5$ from now on, so that the full collection of interaction vertices is $g_1,g_2,g_3,g_\parallel$, and $g_5$.

\subsection{Flow Equations}

Using the definitions of the propagators and vertices of the above sections, one can directly derive the flow equations for the model. The result is
\begin{eqnarray}
\dot{\Sigma}_{sSC} = \frac{1}{2}S_{sSC}(g_{1} + g_{2} - 2g_5) \\  \dot{\Sigma}_{CDW} = \frac{1}{2}S_{CDW}(g_{1} + g_3 - g_\parallel)
\end{eqnarray}
for the order parameter flows (off-diagonal self-energies). Here $S_{\mbox{$s$SC}}$($S_{\mbox{CDW}}$) denote loop integrals over the corresponding matrix elements of the single scale propagator defined in the last section.

 For the interaction vertices the flow equations are given by
\begin{eqnarray}
 \dot{g}_{1} &=& -(A+D)g_5^2 -\frac{1}{2}C g_3^2 - D g_3 g_\parallel - \frac{1}{2}C g_\parallel^2 - (C+D)g_{1}^2 - g_{1}(C g_5 + B g_\parallel + A g_{2}) \nonumber\\  & & - g_{2}(C g_5 + \frac{1}{2} D g_{2}) \nonumber\\ 
  \dot{g}_{2}&=& -(A+D)g_5^2 - \frac{1}{2}(B + C)g_3^2 - C g_\parallel^2 - C g_{1}g_5 - \frac{1}{2}A g_{1}^2 - g_{2}(C g_5 + D g_3 - \nonumber\\  & & D g_\parallel + D g_{1}) - \frac{1}{2}(A+B+2C)g_{2}^2 \nonumber
 \\ \dot{g}_\parallel &=& -\frac{1}{2}B g_3^2 + (2D - \frac{1}{2}A - \frac{1}{2}B)g_\parallel^2 - D g_3g_{1}^2 - \frac{1}{2}B g_{1} - (C g_5 + D g_3 - D g_\parallel + D g_{1})g_{2} \nonumber\\ & & - \frac{1}{2}(A+B+2C)g_{2}^2 \nonumber\\ 
 \dot{g}_3 &=& -D g_3^2 - g_3(B g_\parallel + C g_{1} + (B + C)g_{2}) - D g_\parallel g_{1} - \frac{1}{2}D g_{2}^2 \nonumber\\  \dot{g}_5 &=& \frac{1}{2} C(g_3^2 + g_\parallel^2) - C g_{1} g_{2} - C g_5^2 - (A+D)(g_{1} + g_{2})g_5.
 \end{eqnarray}
The capital letters are abbreviations for the loops occuring on the RHS of the flow equations. They are explicitly given by
\begin{eqnarray*}
A =& \frac{1}{2}\partial_\chi T\sum_\omega\int dk \mathcal{G}_{11}(\omega,k) \mathcal{G}_{33}(\omega,k) &\qquad \mbox{particle-hole loop} \\
B =& \frac{1}{2}\partial_\chi T\sum_\omega\int dk \mathcal{G}_{11}(\omega,k) \mathcal{G}_{33}(-\omega,-k) &\qquad \mbox{particle-particle loop} \\
C =& \frac{1}{2}\partial_\chi T\sum_\omega\int dk \left(\mathcal{G}_{21}(\omega,k)\right)^2 &\qquad \mbox{$s$SC loop} \\
D =& \frac{1}{2}\partial_\chi T\sum_\omega\int dk \left(\mathcal{G}_{31}(\omega,k)\right)^2 &\qquad \mbox{CDW loop}.
\end{eqnarray*}
The Matsubara sums appearing in the loops may be evaluated analytically, whereas the momentum integrals have to be evaluated numerically in general. However, at $T=0$ and with a linearized dispersion, all integrals can be obtained analytically, although they are quite complicated. \\
Finally, following Gersch et al. \cite{gersch}, we also calculate the free energy of the system, which is essentially determined by the flow of the zero-point 1PI vertex and the free energy of the initial, non-interacting, sytem. The equation can be written as\cite{gersch}
\begin{equation}
\dot \Omega = - \frac{1}{2\chi} \Tr\left(\Sigma\mathcal{G}\right),
\end{equation}
where the intial condition $\Omega (\chi=0)=\Omega_0$ is calculated by integrating out the non-interacting fermions at $\chi=0$. Here of course the symmetry breaking terms in the quadratic part of the action have to be taken into account.

\subsection{Self-Consistent Approach: Results Depend on the Choice of Counterterm}

\begin{figure}[h]
\centering
\includegraphics[width = 6cm]{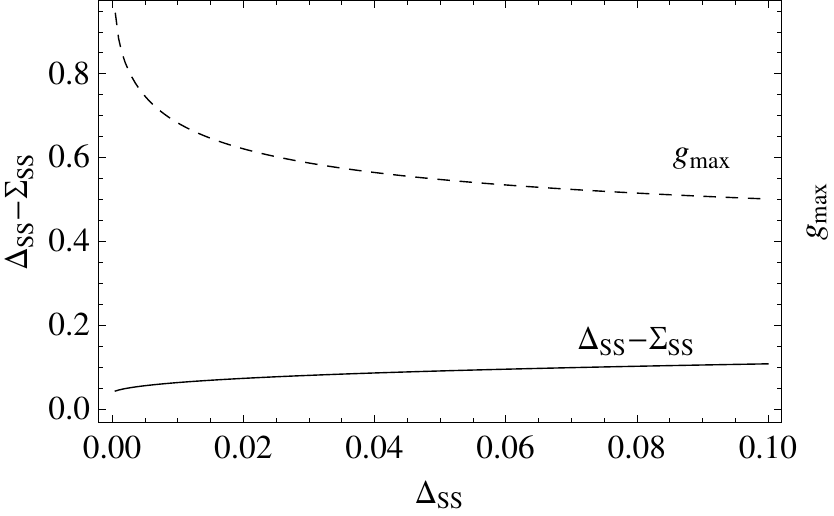}
\label{fig:sod}
\caption{Counterterm dependence of the $s$SC gap $\Delta_{SS}- \Sigma_{SS}$ deep in the $s$SC regime (solid line). Also shown is the largest coupling $g_{max}$ encountered during the flow (dashed line).}
\end{figure}

 When the truncation of the flow equations is not exact (as is the case for generic models), the results of the fRG with counterterms will in general no longer be counterterm independent. In fact, this is what happens for our example of the Tomonaga-Luttinger model. The problem occurs even if one investigates regions of the phase diagram where only one instability is dominant. 
This is shown in Fig. \ref{fig:sod} for the case of dominant $s$SC instabiltiy for temperatures the critical temperature $T_c$ where symmetry breaking sets in. In other words, the selfenergies at the end of the flow become function of the counterterms $\Delta$. Hence we write $\Sigma (\chi,\Delta)$.
One also finds that for low temperatures not all choices of counterterm lead to convergent flows.

  Thus a method is needed to choose one value of the counterterm for which the results are expected to be optimal in some sense. We will follow an analogy to perturbation theory, or more precisly a method for determining optimal counterterms based on a variant of renormalized perturbation theory. This was used by Neumayr and Metzner \cite{neumayr} to circumvent the occurence of divergencies in the renormalized perturbation expansion of many-fermion systems at $T=0$. The problem there is that terms like
\begin{displaymath}
\mathcal{G}_0\Sigma\cdots\mathcal{G}_0\Sigma\mathcal{G}_0
\end{displaymath}
occur in the expansion which lead to divergencies at the Fermi surface because of the singular behaviour of the Green's functions. 
 The strategy followed by Neumayr and Metzner is to choose a different starting point for the perturbation expansion by introducting a counterterm $\delta\ham$ with
\begin{eqnarray}
\tilde{\ham}_0 = \ham_0 + \delta\ham \nonumber\\
\tilde{\ham}_I = \ham_I - \delta\ham,
\label{deltaham}
\end{eqnarray}
and to choose $\delta\ham$ such that the self-energy vanishes everywhere on the Fermi surface, i.e.
\begin{equation}
\tilde{\Sigma}(k)\large{ |}_{\epsilon_k = 0} = 0.
\label{scpert}
\end{equation}
This equation is solved iteratively to the desired order in perturbation theory, thus enabling one to perform calculations that remain finite as the zeroes of $\tilde{\Sigma}$ cancel the poles of $\mathcal{G}_0$ at the Fermi surface. 

 The physical idea behind this method is to choose an optimal starting point for the perturbative expansion, that is one where the non-interacting part already describes the system as well as possible. It is very natural to assume that the quality of a starting point - i.e. a choice of $\ham_0$ - is measured by the amount that $\ham_0$ is renormalized. If it is not renormalized at all, which is ensured for low energies by eq.(\ref{scpert}), it is obviously already a good description of the full interacting model. 

 The connection to the fRG approach to broken symmetries is established by noting that the method still works for symmetry breaking self-energies, such as superconductivity \cite{neumayr}. Furthermore, in the fRG scheme with counterterms one makes exactly the same division of the system into an interacting and a free part as in Eq. \ref{deltaham}. As the truncation of the flow equations corresponds to a weak-coupling approximation, one is also interested in having a good starting point for the perturbation expansion, so that higher order corrections can be expected to be small.

Based on these similarities, we propose to choose the counterterm $\Delta$ using the same condition as in eq.(\ref{scpert}), which in fRG terms reads
\begin{equation}
\Sigma(\chi = 1, \Delta) = 0. \label{self}
\end{equation}
This equation is understood to hold for selfenergy appearing in the interacting Green's function as $\chi \Sigma$ with initial condition $\Sigma (\chi=0, \Delta)=  \Delta$. It does not correspond to a trivial situation, as the total or effective selfenergy, i.e. the difference to the non-interacting quadratic part, at $\chi=1$ is given by $\Sigma - \Delta$.

\begin{figure}[h]
\centering
\label{convergence}
\includegraphics[width = 8cm]{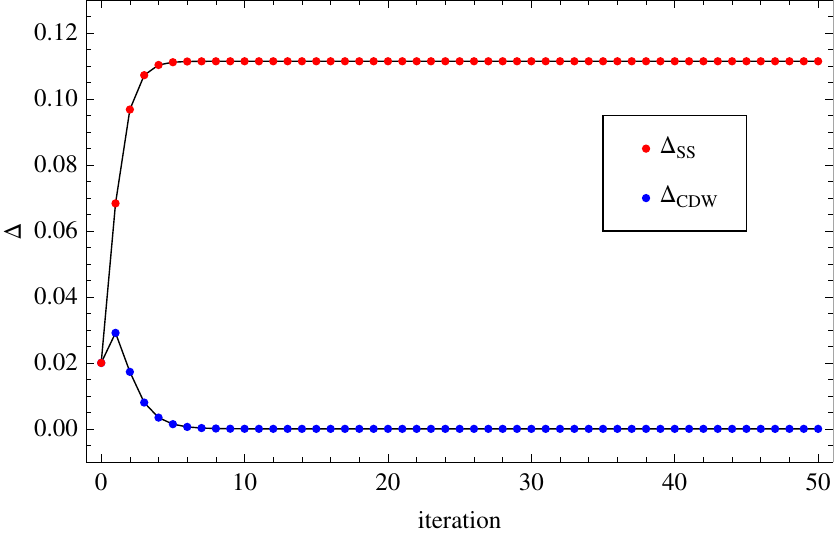}
\includegraphics[width = 8cm]{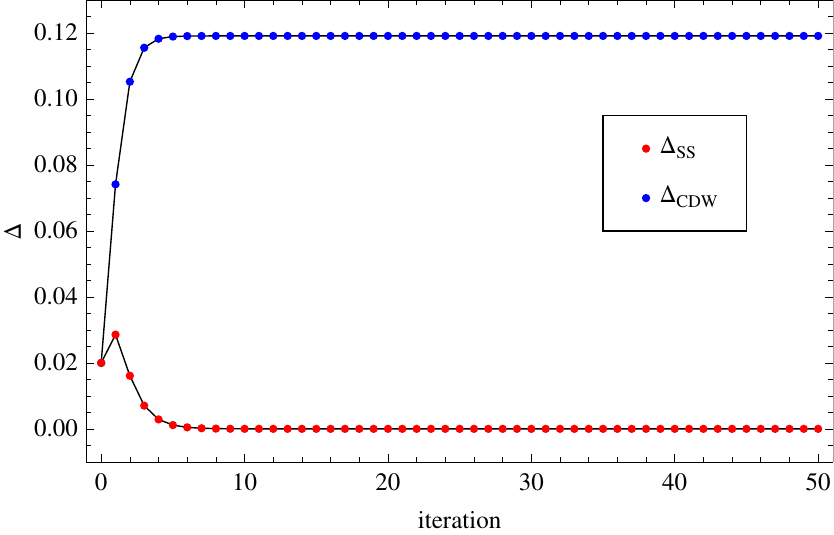}
\includegraphics[width = 8cm]{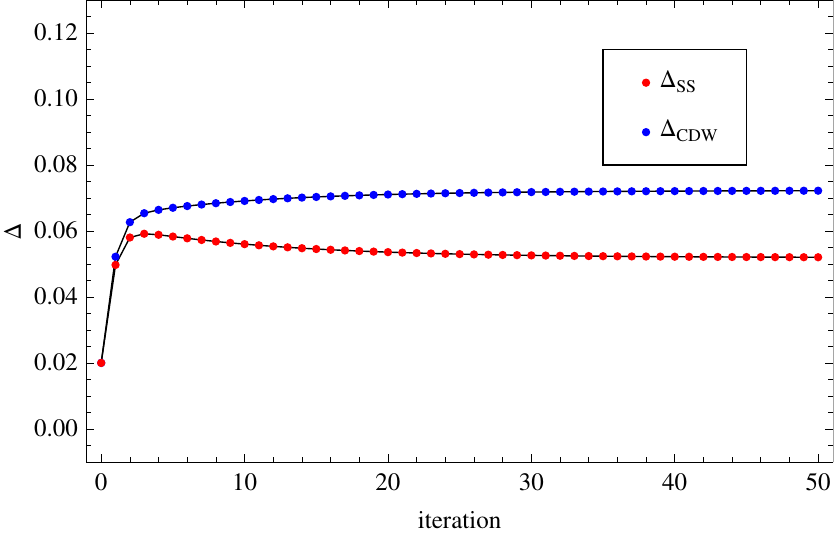}
\caption{Convergence of iterative solution of the self-consistency condition for $g_1=-0.3$, $g_3 = -0.15$, in units where $v_F=1$,  a) in the $s$SC regime ($g_2=0$), b) in the CDW regime ($g_2=-0.5$), and c) close to the transition ($g_2=-0.25$).}
\end{figure}

Of course, Eq. \ref{self} again can only be solved iteratively by inserting the final results for the negative total selfenergy $-(\Sigma - \Delta)$ at $\chi=1$ as counterterm into a new run.  However, as shown in Fig. \ref{convergence}, the iteration converges very quickly to a unique value. Thus it is possible to make a well-defined choice of counterterm out of a continuum of possibilities at moderate computational costs. In the case of the Tomonaga-Luttinger model considered here, typically less than ten iterations were needed for the method to converge.

\subsection{fRG Results} \label{sec:cdw}
\begin{figure}
\centering
 \includegraphics[width=12cm]{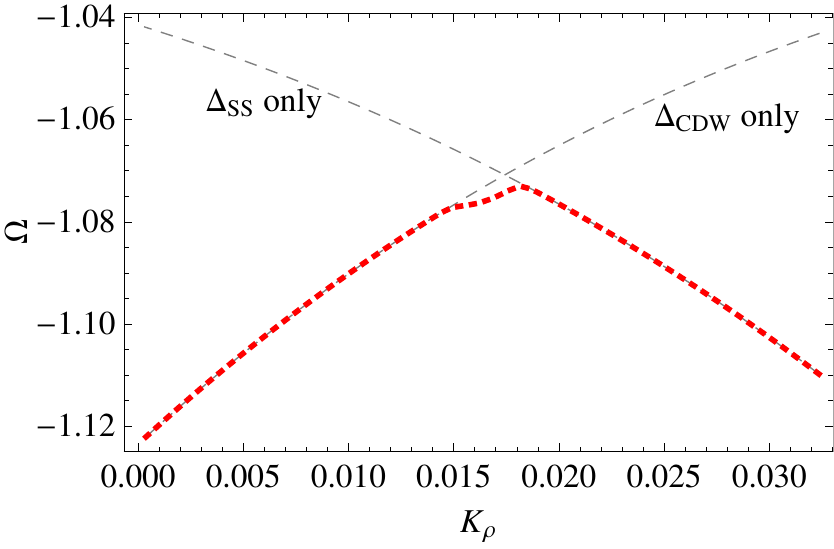}
 \label{fig:omega}

\caption{Free energy for flows using only one counterterm (CDW or $s$SC) at a time (gray lines) and for flows with both counterterms (red line) in the vicinity of the phase transition. $g_3 = - 0.5$ is kept fixed whilst $K_\rho$ is varied.}
\end{figure}

\begin{figure}
\centering
\includegraphics[width=12cm]{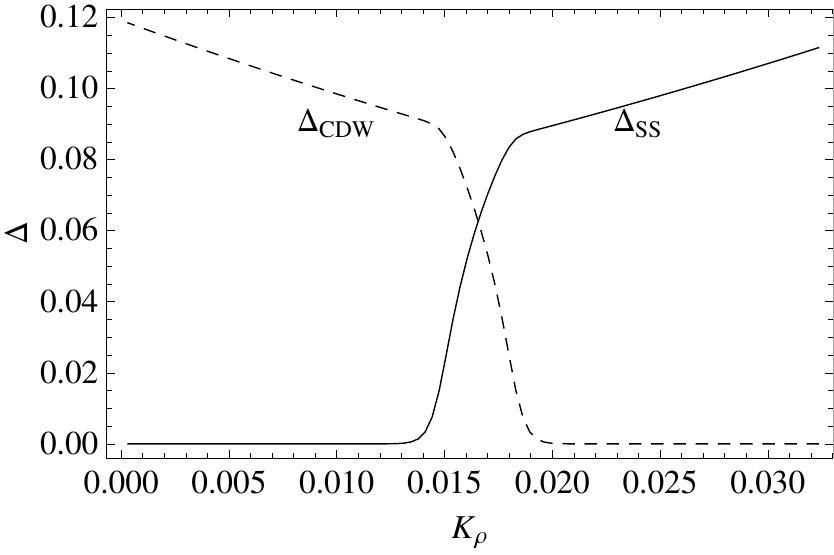}
\label{fig:gaps}
\caption{Behavior of the CDW and $s$SC order parameters in the vicinity of the phase transition. $g_3 = - 0.5$ is kept fixed whilst $K_\rho$ is varied, for flows with both countertems included.}  
\end{figure}

As discussed before, for $g_3<0$, the bosonization description yields a quantum phase transition (QPT) in the $K_\rho-g_3$-plane between the power-law superconductor and the charge-density wave ground state. The question is wether the fRG scheme can reproduce this QPT. Hence we run the fRG as described above for a range of system parameters near the QPT. We use two criteria for determining the position of the phase transition: First, we use either the CDW or the $s$SC counterterm separately, and also calculate the free energy during the flow. We then choose the solution with lower energy (as done in Gersch et al.\cite{gersch}) as the best approximation to the true behaviour. The resulting free energies are shown in Fig. \ref{fig:omega} for $g_3=-0.5$ (in units where the lattice constant is unity and $v_F=1$). We can read off a critical value of $K_\rho \approx 0.017$ where the energetically favorable order parameter changes from pairing to CDW. 
Second, we can use both counterterms simultaneously, allowing thus for a coexistence of both orders, and say that for $\Delta_{\text{CDW}}(\chi=1) > \Delta_{\text{SC}}(\chi=1)$ the system is in the CDW phase and vice versa. In this case, the free energy for flows with order parameters allowed does only differ a little from the curves for only one order parameter in  a small coexistence region between $0.015$ and $0.019$ (see  Fig. \ref{fig:omega}). 
The superconducting order parameter suppresses the CDW order on the large-$K_\rho$-side, while the situation is reversed at small $K_\rho$.
This behavior of the order parameters versus $K_\rho$ is shown in Fig. \ref{fig:gaps}. Again we get a transition point of $K_\rho \approx 0.017$. 
Hence, we find that the results obtained by these two ways to determine the correct counterterms (via order parameters and via free energies) coincide in the selection of the dominant order parameter. 

\begin{figure}
 \centering
\includegraphics[width=12cm]{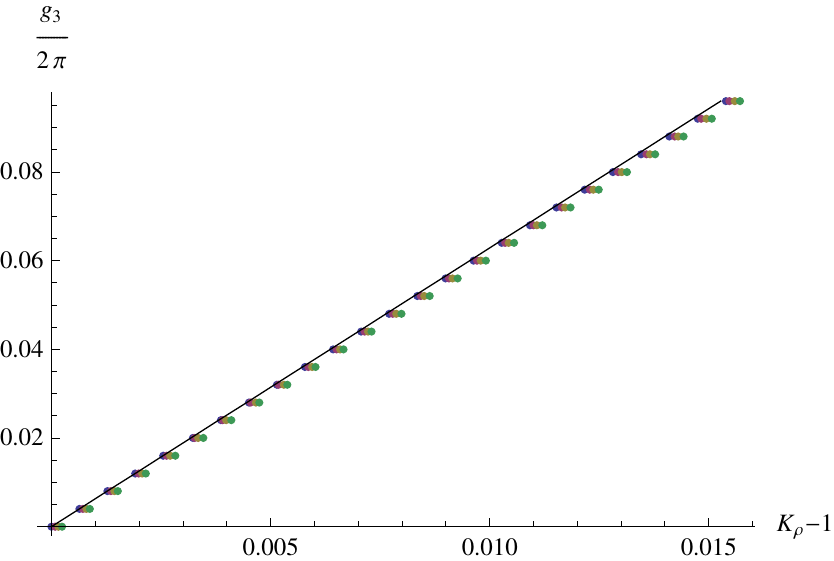}
\label{fig:g1dependence}
\caption{Phase diagram in the $K_\rho-g_3-$plane for different values of $g_1$. The bosonization result (solid line) is independent of $g_1$, while in the fermionic RG calculation there is a slight dependence on its value (colored points, $g_1 = 0.5$, $1$, $1.5$, $2$).}
\end{figure}

Next, as a further check of the reliability of the result, we use the fact that in the bosonization method the spin and charge sectors decouple, so that the phase diagram is independent of the value of $g_1$ (as long as the spin sector is in the massive phase), and only depends on the combination $g_\parallel - g_2$ entering $K_\rho$. 
We find that even though this separation into different sectors is not manifest in the fermionic formulation, the position of the phase transition changes only slightly if $g_1$ is changed with $K_\rho$ and $g_3$ fixed. This is shown in Fig. \ref{fig:g1dependence}. All this evidence taken together demonstrates that the fermionic fRG with counterterms is able to handle systems with more than one instability in a satisfactory way.

On the back side it should be mentioned that near the transition between SC and CDW regime the fRG predicts a regime with coexisting order, where both order parameters are nonzero at the end of the flow. This regime is absent in the bosonization, where the incompressible CDW has exponentially decaying charge correlations. The fRG results do never give a gapless fermionic spectrum and replace the quantum critical point by a small coexistence region.
While this difference is presumably due to the approximate treatment of the competition between the two ordering tendencies via static order only, it is certainly not so severe as to render the fRG approach useless. The ultimate nature of quantum phase transitions is often a subtle issue.
Furthermore, as a more severe but also understandable difference of the fRG picture to the bosonization, the power-law superconductor is replaced by a long-range ordered symmetry-broken state. This was alluded already in the beginning of this section and is clearly due to our approximation in which only frequency-independent selfenergies are allowed, and where the simplified wave-vector- and frequency-structure of the vertices excludes a correct description of collective fluctuations.  

\begin{figure}[tb]
\centering
\includegraphics[width=7cm]{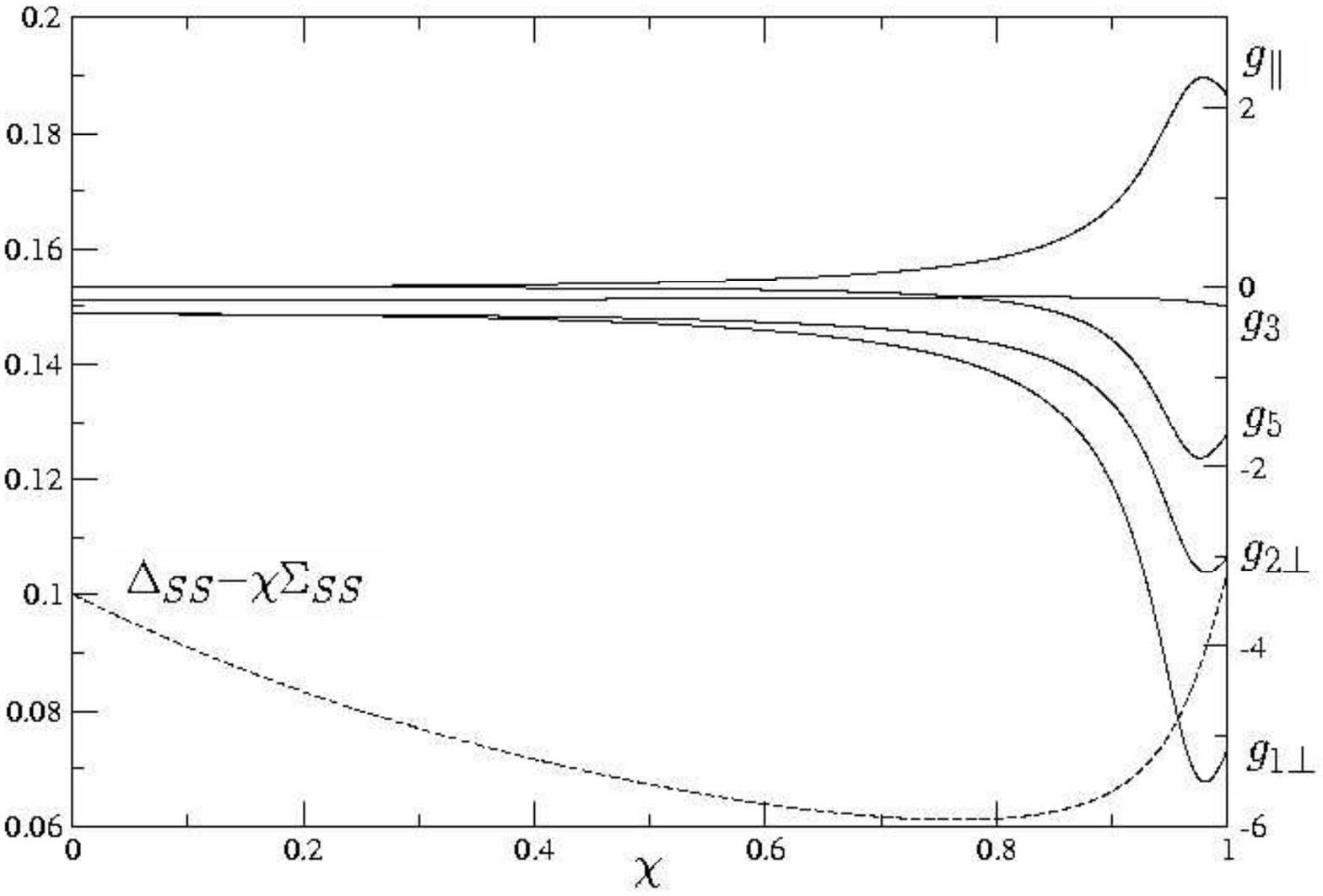}
\includegraphics[width=7cm]{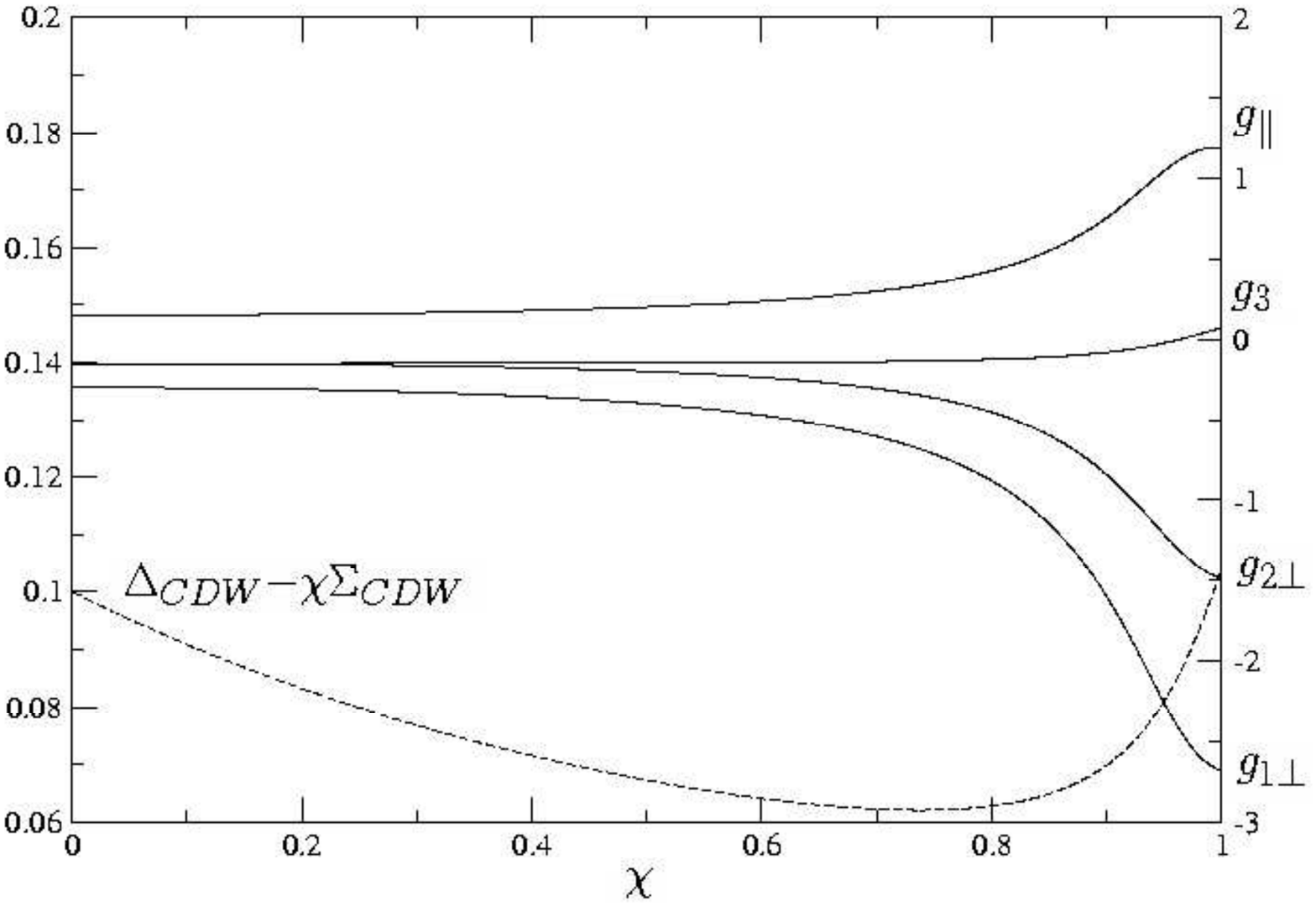}
\caption{Typical fRG flows a) in the massless (SS) phase and b) in the massive (CDW) phase. The scale for the order parameters is on the left side, the scale for the couplings (multiplied by $-1$) on the right side.} 
\label{fig:typflows}
\end{figure}

In Fig. \ref{fig:typflows} we show typical flows in the $s$SC and the CDW regime. In both regimes, all couplings remain moderate, such that the truncation error should remain acceptable. 
Couplings in the massless phase, where superconductivity dominates, become larger due to the presence of the Goldstone mode \cite{lauscher}. In an exact treatment one would actually observe a divergence of the couplings which reflects the fact that the Goldstone mode is massless. In the approximate treatment, this seems to be violated to some extent, i.e. the final $s$SC order parameter may be too large compared to the unknown exact result. In other words, the bias or explicit symmetry breaking of the initial conditions has not been removed entirely in the approximate flow. Of course one could try to vary the counterterm so as to recover the Goldstone mode. However it is unclear if this scheme would actually work, and moreover this alternative scheme for selecting the right counterterm could not be generalized to discrete symmetry breakings. We believe that for our purposes the imprecise treatment of the Goldstone mode is a quantitative problem that can be reduced by more sophisticated treatments of the frequency and momentum dependence of the vertex functions. Due to the qualitative correctness of the transition line between dominant CDW and superconducting correlations, the counterterm method in its present form should still be to determine the main picture.

\section{Conclusions and Outlook}
Summarizing the above results, we have investigated the applicability of the interaction (flat cutoff) fRG flow scheme using counterterms to situations with symmetry breaking and competing orders. 
We have found that this method can handle problems where there is only on channel causing instabilities (the reduced BCS model) perfectly well. 
We then went on to apply the scheme to a system with competing instabilities, which led to several new insights. 
First of all, the results at the end of the flow are no longer independent of the choice of counterterm, so that one has to find a way to choose a preferred value of counterterm. 
We proposed to use solve a self-consistency equation iteratively, which turned out to give unique and sensible results in most cases. 
With this device at hand we investigated  the quantum phase transition from a CDW ordered state to state with (power-law) superconducting order in a one-dimensional model with attractive fermions at half band filling. 
Two different ways to run the flows, either  with only one type of symmetry breaking allowed, or with both types of symmetry breakings included gave the same location of the transition line from dominant CDW to dominant superconducting correlations, in agreement with the bosonization result. For most parameters, the coupling constants remain in the weak coupling range.

Expected differences with respect to bosonization occurred in the long-range-ordered nature of the superconducting ground state in the fRG, and in the small coexistence region near the transition, smearing out the quantum criticality. Improvements in this regard would require a more precise treatment of collective fluctuations, which is beyond the scope of this work. Nevertheless, the present method is able to predict correctly the type of leading ground state correlation and to produce a gapped renormalized single-particle spectrum. All fermionic modes can be integrated out without a severe run-away flow of the couplings. 
As a next step, the scheme should be applied to two-dimensional models. There it should allow one to compute, e.g., the order parameter magnitude and the gap structure around the Fermi surface of unconventional superconductors.

We thank Manfred Salmhofer, Walter Metzner, Jutta Ortloff, T. Maurice Rice and Manfred Sigrist for useful discussions.  This work was supported by the DFG research unit FOR723.


\begin{thebibliography}{99}
\bibitem{wilson} K. Wilson, Phys. Rev. B \textbf{4}, p. 3174 (1971); Phys. Rev. B \textbf{4}, p. 3184 (1971).
\bibitem{polchinski} J. Polchinski, Nucl. Phys. B \textbf{231}, p. 269 (1984).
\bibitem{wetterich} C. Wetterich, Phys. Lett. B \textbf{301}, p. 90 (1993).
\bibitem{tnt}  M. Salmhofer, C. Honerkamp, Prog. Theor. Phys. \textbf{105}, 1 (2001).

\bibitem{goettingen} V. Meden, W. Metzner, U. Schollw\"ock, and K. Sch\"onhammer, Phys. Rev. B {\bf 65}, 045318 (2002).
\bibitem{kopietz} P. Kopietz, T. Busche, Phys. Rev. B {\bf 64}, 155101 (2001).

\bibitem{zanchi} D. Zanchi, H. H. Schulz, Phys. Rev. B \textbf{67}, 5909 (1996).
\bibitem{halboth} C. J. Halboth, W. Metzner, Phys. Rev. B {\bf 61}, 7364 (2000); Phys. Rev. Lett. {\bf 85}, 5162 (2000).
\bibitem{rice} C. Honerkamp, M. Salmhofer, N. Furukawa, T. M. Rice, Phys. Rev B \textbf{63}, 035109 (2001).

\bibitem{katanin} A. P. Kampf and A. A. Katanin, Phys. Rev. B {\bf 67}, 125104 (2003); Phys. Rev. Lett. {\bf 93}, 106406 (2004).

\bibitem{tempflow} C. Honerkamp, M. Salmhofer, Phys. Rev. B \textbf{64} 184516 (2001).

\bibitem{wang} F. Wang, H. Zhai, Y. Ran, A. Vishwanath, D.-H. Lee, Phys. Rev. Lett. {\bf 102}, 047005 (2009).
\bibitem{lauscher} M. Salmhofer, C. Honerkamp, W. Metzner, O. Lauscher, Prog. Theor. Phys. {\bf 112},  942 (2004).

\bibitem{gersch} R. Gersch, J. Reiss, C. Honerkamp, New J. Phys. \textbf{8} 320 (2006).

\bibitem{gersch2d} R. Gersch, C. Honerkamp, W. Metzner, New J. Phys. {\bf 10}, 045003 (2008).

\bibitem{bosonizerefs} T. Baier, E. Bick and C. Wetterich, Phys. Rev. B {\bf  70}, 125111
(2004); H. C. Krahl, J. A. M\"uller, C. Wetterich, Phys. Rev. B {\bf 79}, 094526 (2009); P. Strack, R. Gersch, W. Metzner, Phys. Rev. B {\bf  78}, 014522 (2008); L. Bartosch, P. Kopietz, A. Ferraz, Phys. Rev. B {\bf 80}, 104514 (2009).
\bibitem{gies} H. Gies and C. Wetterich, Phys. Rev. {\bf D 65}, 065001 (2002).
\bibitem{peskin} M. Peskin, D. Schroeder, \emph{An Introduction to Quantum Field Theory}, Addison-Wesley (1995).

\bibitem{altland} A. Altland, B. Simons, \emph{Condensed Matter Field Theory}, Cambridge University Press (2006).
\bibitem{kataninmod} A. A. Katanin, Phys. Rev. B \textbf{70}, 115109 (2004).

\bibitem{intflow} C. Honerkamp, D. Rohe, S. Andergassen, T. Enss,  Phys. Rev. B \textbf{70},  235115 (2004)

\bibitem{neumayr} A. Neumayr, W. Metzner, Phys. Rev. B \textbf{67}, 035112 (2003).

\bibitem{senechal} D. S\'{e}n\'{e}chal, \emph{An Introduction to Bosonization}, cond-mat/9908262.

\bibitem{giamarchi} T. Giamarchi,\emph{Quantum Physics in One Dimension}, Oxford University Press (2004).
\bibitem{rajaraman} R. Rajaraman, \emph{Solitons and Instantons: An Introduction to Solitons and Instantons in Quantum Field Theory}, North Holland (1982).


\bibitem{nakahara} M. Nakahara, \emph{Geometry, Topology and Physics}, IoP Publishing (2003).









\end{thebibliography}
\end{document}